\documentclass[twocolumn,showpacs,preprintnumbers,amsmath,amssymb]{revtex4}
\usepackage{graphicx}% Include figure files
\usepackage{dcolumn}% Align table columns on decimal point
\usepackage{bm}% bold math

\def\bea{\begin{eqnarray}}
\def\eea{\end{eqnarray}}

\begin{document} 

\preprint{Version 2.4}

\title{Azimuth quadrupole component spectra on transverse rapidity ${\bf y_t}$ \\ for identified hadrons from Au-Au collisions at $\sqrt{s_{NN}} =$ 200 GeV}

\author{Thomas A. Trainor}
\address{CENPA 354290, University of Washington, Seattle, WA 98195}

%%%%%%%%%%%%%%%%%%%%%%%%%%%%%%%

\date{\today}

\begin{abstract}
I present the first isolation of azimuth quadrupole components from published $v_2(p_t)$ data (called elliptic flow) as spectra on transverse rapidity $y_t$ for identified pions, kaons and Lambdas/protons from minimum-bias Au-Au collisions at 200 GeV. The form of the spectra on $y_t$ indicates that the three hadron species are emitted from a common boosted source with boost $\Delta y_{t0} \sim 0.6$. The quadrupole spectra have a L\'evy form similar to the soft component of the single-particle spectrum, but with significantly reduced ($\sim 0.7\times$) slope parameters $T$. Comparison of quadrupole spectra with single-particle spectra suggests that the quadrupole component comprises a small fraction ($< 5$\%) of the total hadron yield, contradicting the hydrodynamic picture of a thermalized, flowing bulk medium. The form of $v_2(p_t)$ is, within a constant factor, the product of $p'_t$ ($p_t$ in the boost frame) times the ratio of quadrupole spectrum to single-particle spectrum. That ratio in turn implies that above 0.5 GeV/c the form of $v_2(p_t)$ is dominated by the hard component of the single-particle spectrum (interpreted as due to minijets). It is therefore unlikely that so-called {\em constituent-quark scaling} attributed to $v_2$ is relevant to soft hadron production mechanisms (e.g., chemical freezeout).
\end{abstract}


\maketitle

%%%%%%%%%
 \section{Introduction}

Measurement and interpretation of ``elliptic flow''  ($v_2$) is considered of central importance to the RHIC program because it provides the main support for interpretation of RHIC collisions as producing sQGP---a thermalized {\em strongly-coupled} partonic medium with very low viscosity, sometimes described as a ``perfect liquid''~\cite{teaney,hirano2,csernai}. The differential form $v_2(p_t)$ in particular is a keystone of that interpretation~\cite{keystone,teaney2,kolb3,huov2}. 

In the conventional flow description $v_2(p_t)$ for identified hadrons follows trends predicted by hydrodynamic (hydro) models at smaller $p_t$~\cite{kolb1,kolb2,huov2} but ``saturates'' at larger $p_t$ where parton fragmentation is expected to dominate~\cite{v2sat}. $v_2(p_t)$ ``scaling'' is used to demonstrate that ``constituent quarks'' dominate hadronization. $v_2$ and $p_t$ divided by constituent quark number $n_q$ (i.e., 2 for mesons, 3 for baryons) appear to be related by a universal curve, the inference being that hadrons are formed by quark coalescence from a thermalized partonic medium~\cite{quarkcoal}. Further evidence for collective {\em partonic} flow is inferred from $v_2$ data for selected hadrons such as the $\phi$ and $D$ mesons and $\Xi$ and $\Omega$ baryons, where elliptic flow generated by {\em hadronic} rescattering should be small~\cite{0701010}.

The large $v_2$ at RHIC energies, described in a hydro context as elliptic flow, is thus interpreted to imply early thermalization of a collective partonic medium resulting in large pressure gradients which drive the development (for an azimuthally asymmetric system) of the observed azimuth eccentricity in hadronic momentum space~\cite{huov2}.

Logical, technical and interpretational problems have emerged for $v_2$. Formation of a thermalized partonic medium may imply measured $v_2$ systematics in a hydro context, but do $v_2$ data {\em require} the hydro interpretation? If more accurate measurements of $v_2$ or a different asymmetry measure are introduced can the hydro interpretation be falsified? Is hadron formation from an extended QCD field system in some sense a {\em universal} characteristic of all nuclear collisions at RHIC energies, including N-N collisions? Is partonic or hadronic rescattering {\em required} to produce a system which {appears} to be thermalized?

$v_2(p_t)$ as defined is a ratio of two spectra, confusing single-particle two-component physics with the physics of the {\em azimuth quadrupole} (cf.~\cite{flowmeth} for quadrupole terminology). $v_2$ data are therefore difficult to interpret directly in terms of conventional spectrum analysis. Comparisons are typically made indirectly {\em via} hydrodynamic models whose validity can be questioned, especially because they do not model important aspects of single-particle spectra.

Recent initiatives have shed new light on the azimuth quadrupole problem. Single-particle spectra for identified hadrons have been accurately separated into soft and hard components (longitudinal and transverse fragmentation)~\cite{2comp}. No evidence for {collective} radial flow was found. An analysis of the algebraic structure of $v_2$ and alternative measures reveals that two-particle correlations are basic to {\em any} $v_2$ measurement, and $\eta$ dependence of 2D angular autocorrelations can be used to isolate azimuth quadrupole correlations from ``nonflow''~\cite{flowmeth,gluequad}. Re-examination of the centrality dependence of published $p_t$-integrated $v_2$ data reveals a simple dependence on the number of binary collisions, and minijets are identified as the dominant source of nonflow~\cite{gluequad}.

In this analysis spectra associated with the azimuth quadrupole are extracted from published $v_2(p_t)$ data and plotted on transverse rapidity $y_t$. Those spectra determine the quadrupole source boost and relative hadron abundances. The main goals of the present analysis are to identify the quadrupole component as single-particle spectra for each hadron species and to determine the abundance of quadrupole hadrons for each species. 

Quadrupole spectra and associated hadron yields could play a critical role in tests of hydro model validity and claims of ``perfect liquid'' in response to open questions. For instance, what are the spectrum properties of the quadrupole, and what fraction of all produced hadrons does the quadrupole component represent? I.e., do almost all particles in a collision participate in elliptic flow as widely assumed (a {\em truly collective} flowing bulk medium), or is the quadrupole component an isolated process involving a small fraction of the total system? Specific tests of hydro theory are reserved for subsequent analysis.

% \subsection{Plan}

This paper is arranged as follows. The analysis procedure is briefly outlined. New methods derived from two-particle correlation analysis are reviewed, and conventional elliptic flow analysis is interpreted in that broader context. A sample of $v_2(p_t)$ data for three hadron species is transformed between different plotting formats to illustrate the requirements for a full analysis and its likely outcome. The full analysis includes three steps: 1) specify a two-component representation of the single-particle spectrum for each hadron species, 2) incorporate the kinematics of boosted particle sources, 3) extract and interpret azimuth quadrupole spectra.

Steps 1), 2) and 3) are used to obtain quadrupole spectra on {\em transverse rapidity} $y_t$ from selected $v_2(p_t)$ data. Quadrupole spectrum shapes are compared to single-particle spectra to search for quadrupole manifestations therein and to place limits on quadrupole absolute yields. Quadrupole and minijet contributions to spectra are compared to determine the relations between them, and several $v_2(p_t)$ scaling relations and their implications for claims of sQGP are examined in the context of this analysis.

%%%%%%%%%%%
\section{Analysis description}

Conventional differential flow measure $v_2(p_t)$ as defined includes a ratio of two hadron spectra: the spectrum of ``flowing particles'' (quadrupole component) in the numerator and the azimuth-averaged single-particle spectrum in the denominator. Although arguments from a hydrodynamic context favor ratio $v_2$, it is important to examine the quadrupole spectrum (numerator) directly. 

The goal of this analysis is to isolate from existing $v_2(p_t)$ data the spectra on $y_t$ of hadrons associated with the quadrupole component for three hadron species, and to compare quadrupole spectra with azimuth-averaged single-particle spectra. The analysis should reveal the radial boost distribution of the particle source and the fractional yields of quadrupole hadrons in a collision. The analysis should improve our understanding of the underlying physical mechanism. E.g.,  is it hydrodynamic expansion~\cite{schned} or QCD field interactions~\cite{gluequad}?

%\subsection{Analysis method}

A hint of the benefits of this analysis is obtained by plotting $v_2(p_t) / p_t(\text{lab frame})$ {\em vs} the proper transverse rapidity $y_t$ for each hadron species, as in Sec.~\ref{v2meas}. To understand why that strategy provides qualitatively new information the kinematics of boosted thermal sources are reviewed in Sec.~\ref{boost}. To isolate the quadrupole spectrum (numerator) from the $v_2(p_t)$ ratio corresponding single-particle spectra (denominator) are presented in Sec.~\ref{singlespec}. 

In the full analysis Fourier amplitude $V_2(p_t)$ is recovered from $v_2(p_t)$ by eliminating the single-particle spectrum from its denominator.  Based on the Cooper-Frye description of a thermal source boosted on transverse rapidity $y_t$ a factor $p'_t$ ($p_t$ in the boost frame) is also removed to form an approximate expression for quadrupole spectrum $\rho_2(y_t;\Delta y_{t0})$.  There remains an $O(1)$ factor due to an integral approximation and ambiguity between the quadrupole boost $\Delta y_{t2}$ and the absolute yield of the quadrupole spectrum $n_{ch2}$. Comparisons between quadrupole and single-particle spectra, including the hard component (scattered-parton fragments), constrain the absolute quadrupole spectrum and yield.

%%%%%%%%%
\section{Azimuth Correlation Analysis}

Two-particle azimuth correlation analysis is outlined, and conventional differential (on $p_t$) elliptic flow analysis is described in the larger context.  Given a $p_t$ spectrum defined in histogram form with bins of width $\delta p_t$, the symbol $v_2(p_t)$ indicates the value of $v_2$ in a $p_t$ bin with bin multiplicity $n_{p_t}$. The measured integral quantities for each collision event of $n$ particles are vector Fourier coefficients $\vec Q_m = \sum_{i=1}^n \vec u(m\phi_i)$ and scalar {\em power-spectrum} elements $V_m^2 = \sum_{i \neq j=1}^{n,n-1} \vec u(m\phi_i)\cdot \vec u(m \phi_j)$~\cite{flowmeth,gluequad}.  The same quantities in differential form can be defined as 1D and 2D histograms respectively on $p_t$ bins.

\subsection{Two-particle correlations on ${\bf p_t}$}

Two-particle azimuth correlations can be studied without introducing a reaction or event plane. The basic measures of sinusoidal azimuth correlations are the Fourier power spectrum elements $V_m^2$~\cite{flowmeth}. The 2D $p_t$-integrated quadrupole term $V_2^2$ can be generalized to a $p_t$-differential form with unit vectors $\vec u(2 \phi_i)$.
\bea 
V_2^2(p_{t1},p_{t2}) &\equiv& \sum_{i\in p_{t1}\neq j\in p_{t2}=1}^{n_{p_{t1}},n_{p_{t2}}} \cos( 2[\phi_i -\phi_j]) \\ \nonumber
&=& \sum_{i\in p_{t1}\neq j\in p_{t2}=1}^{n_{p_{t1}},n_{p_{t2}}} \vec u( 2 \phi_i) \cdot \vec u(2\phi_j) \\ \nonumber
&\equiv& \vec V_2(p_{t1}) \cdot \vec V_2(p_{t2}),
%\\ \nonumber
%&=& V_2(p_{t1})\, V_2(p_{t2})\, r(p_{t1},p_{t2}),
\eea
where e.g. index $p_{t1}$ labels a histogram bin of nominal width $\delta p_t$ with center at $p_{t1}$ containing $n_{p_{t1}}$ particles. The dot product in the last line defines a mnemonic representation of the $i \neq j$ double sum. Individual {vectors} $\vec V_2(p_t)$ are not accessible. Diagonal element $V^2_2(p_t,p_t)$ denotes the power spectrum element for a single bin centered at $p_t$. 
$V_2^2(p_{t1},p_{t2})$ is an {\em additive} two-particle correlation measure, playing the same role for the azimuth quadrupole that {\em total variance} $\Sigma^2_{p_t:n}$ plays for $p_t$ fluctuations/correlations~\cite{meanptprl}. $V_2^2(p_{t1},p_{t2})$ can describe a two-particle distribution on transverse momentum $(p_{t1},p_{t2})$, mass $(m_{t1},m_{t2})$~\cite{mtxmt} or rapidity $(y_{t1},y_{t2})$.

 %$V_2({p_t}) \equiv  \vec V_2(p_{t}) \cdot \vec V_2$

\subsection{Marginal distribution ${\bf V_2(p_t)}$ {\em vs} ${\bf v_2(p_t)}$}

{\em Marginal} distribution $V_2(p_t)$ is obtained from the asymmetric 2D case that one $p_t$ bin is the entire acceptance (including $n$ particles). I.e., $V_2({p_t})$ is obtained by integrating $V_2^2(p_{t1},p_{t2})$ over one $p_t$ axis
\bea \label{v2dot}
{V_2^2(p_t)} &\equiv& \sum_{i\in p_{t}\neq j=1}^{n_{p_{t}},n-1} \vec u(2\phi_i)\cdot  \vec u(2\phi_j) \\ \nonumber
&=&  \vec V_2(p_{t}) \cdot \vec V_2 \\ \nonumber
V_2(p_t)&= & \frac{\vec V_2(p_{t}) \cdot \vec V_2}{V_2}  \\ \nonumber
 v_2\{2\}(p_t) &\equiv& V_2(p_t) / n_{p_t} \\ \nonumber
&=& \frac{\vec V_2(p_{t}) \cdot \vec V_2}{n_{p_t}\, V_2}.
\eea
The last line defines elliptic flow measure $v_2\{2\}(p_t)$ in terms of two-particle correlations. In general, $V_2(p_t) \neq \sqrt{V^2_2(p_t,p_t)}$. $V_2(p_t)$ is an element of the marginal distribution, whereas $V^2_2(p_t,p_t)$ refers to a single diagonal bin on $(p_{t1},p_{t2})$. The two are related by a covariance.

\subsection{Conventional event-plane method}

Conventional $v_2$ analysis is motivated in the context of an event or reaction plane, but analysis results do not depend on a reaction plane {\em per se}. $v_2$ measures the $m = 2$ Fourier component of {\em any} two-particle azimuth correlations present in collision products, including jet correlations. An ``event plane'' can arise from any event-wise azimuth structure (including minijets), and the ``event-plane resolution'' may not relate to a true reaction plane.  

$p_t$-differential elliptic flow analysis at mid-rapidity is based on a 1D Fourier decomposition on azimuth of the $\eta$-averaged 3D density. The Fourier series is defined in terms of  {\em reaction-plane} angle $\Psi_r$
\bea \label{fourier1}
  \rho(p_t,\phi) = \frac{ V_0}{2\pi} \left\{ 1 + 2\sum_{m=1}^\infty   v_m(p_t) \cos(m[\phi - \Psi_r]) \right\},
 \eea
where $V_0(p_t)/2\pi \equiv \rho_0(p_t)$ is the 3D single-particle $p_t$ spectrum (averaged over $2\pi$ azimuth and one unit of pseudorapidity about $\eta = 0$) described by a {\em two-component} spectrum model~\cite{2comp}. Fourier amplitude {\em ratios} $ v_m(p_t) \equiv  V_m(p_t) /  V_0(p_t) = \langle \cos\{m [\phi - \Psi_r]\}(p_t) \rangle$~\cite{flowmeth}. $V_m$ could represent multiple physical contributions, including {\em minijets} as well as various ``flow'' sources. Eq.~(\ref{fourier1}) is not a conventional Fourier series because common element $\rho_0$ divides each term, thereby coupling all $v_m$. The equation is nonphysical, since $\Psi_r$ is not known {\em a priori}, and the $V_m$ are therefore not measurable by inversion. 

Within the flow model description $\Psi_r$ must be estimated from a subset of the collision products. The estimate is called the {\em event-plane} angle $\Psi_m$, and Eq~(\ref{fourier1}) is rewritten in terms of unit vectors $\vec u(m\phi)$ as
\bea \label{fourier2}
  \rho(p_t,\phi) &=& \frac{1}{2\pi}  \sum_{m = -\infty}^\infty \vec Q_m(p_t)\cdot \vec u(m\phi)  \\ \nonumber
&=& \frac{ Q_0}{2\pi} \left\{ 1 + 2\sum_{m=1}^\infty   q_m(p_t) \cos\left(m[\phi - \Psi_m]\right) \right\},
 \eea
with true {\em Fourier coefficients} $ \vec Q_m(p_t) \equiv \sum_{j \in p_t}^n \vec u(m \phi_j) = Q_m(p_t)\, \vec u(m\, \Psi_m[p_t])$ and Fourier {\em amplitude} {ratios} $q_m(p_t) = Q_m(p_t) / Q_0(p_t)$. The $\vec Q_m$ are conventionally interpreted by assuming that azimuth structure is hydrodynamic in origin (various flows) relating to the reaction plane. However, the $\vec Q_m$ may contain substantial ``nonflow'' contributions dominated by the Fourier coefficients of the same-side minijet peak (jet cone)~\cite{gluequad}. The inferred ``event-plane angle'' $\Psi_m$ (actually the Fourier {\em phase angle}) may be poorly correlated or uncorrelated with the actual A-A reaction plane.

The differential amplitude ratio $q_2(p_t)$ can be obtained by inverting the Fourier series
\bea
q_2(p_t) &=& \langle  \vec u(2 \phi_{i \in p_t}) \cdot \vec u(2 \Psi_2[p_t])\rangle \\ \nonumber
&=&\frac{\vec Q_2(p_t) }{n_{p_t}}\cdot \frac{\vec Q_2(p_t)}{Q_2(p_t)} = \frac{Q_2(p_t)}{n_{p_t}}
\eea
with $n_{p_t} = Q_0(p_t)$. However, according to standard flow-analysis methods $\vec Q_2(p_t) / Q_2(p_t) \rightarrow \vec Q_2 / Q_2 = \vec u(2 \Psi_2)$ which determines the (global?) $m = 2$ event-plane angle, and ``autocorrelations'' (self pairs) must be eliminated from the dot product~\cite{selfpair}. For each particle $i$ in a $p_t$ bin a complementary vector $\vec Q_2 \rightarrow \vec Q_{2i}$ is formed by omitting the $i^\text{th}$  particle from the $\vec Q_2$ sum over $j$. $q_2(p_t)$ then becomes conventional elliptic flow measure $v_{2obs}(p_t)$
\bea \label{v2obs}
v_{2obs}(p_t) &=& \langle  \vec u(2 \phi_{i \in p_t}) \cdot \vec u(2 \Psi_{2i})\rangle \\ \nonumber
&=& \left\langle \frac{\vec Q_2(p_t)}{n_{p_t}}\cdot \frac{\vec Q_{2i}}{Q_{2i}}\right \rangle \\ \nonumber
&\approx& \frac{\vec V_2(p_t)\cdot \vec  V_2}{ n_{p_t}\, \langle Q_{2i}\rangle },
\eea
where the $\vec V_2$ dot product defined in Eq.~(\ref{v2dot}) represents the double sum with $j \neq i$.
$v_{2obs}$ is then divided by the ``event-plane resolution'' $\langle \cos(2[\Psi_2 - \Psi_r])\rangle$ to obtain
\bea  \label{v2ep}
v_2\{EP\}(p_t) &\equiv& \frac{\langle  \vec u(2 \phi_{i \in p_t}) \cdot \vec u(2 \Psi_{2i})\rangle}{\langle \cos(2[\Psi_2 - \Psi_r])\rangle} \\ \nonumber
&=& v_2\{2\}(p_t)\cdot \frac{V_2/\langle Q_{2i}\rangle}{\langle \cos(2[\Psi_2 - \Psi_r])\rangle},
\eea
which gives the exact relation between $v_2\{EP\}$ and $v_2\{2\}$ for the first time in terms of the $O(1)$ second factor~\cite{flowmeth}. The difference between $\{EP\}$ and $\{2\}$ results from a misconception about the $v_{2obs}$ numerator leading to introduction of $\langle Q_{2i}\rangle \sim Q_2$ in the denominator of Eq.~(\ref{v2obs}) in place of $V_2$ as in Eq.~(\ref{v2dot}) (last line)~\cite{flowmeth,gluequad}. The extraneous ``event-plane resolution'' $\sim V_2  / Q_2$ is then introduced to correct $v_{2obs}(p_t)$.

The event-plane method is also described in terms of ``subevents,''~\cite{0409033}. A correlation quantity is defined by
\bea
\langle \vec u(2\phi_{i \in p_t})\cdot  \vec Q_{2i}\rangle&=& \frac{1}{ n_{p_t}} \sum_{i\in p_{t}}^{n_{p_{t}}} \vec u(2\phi_i)\cdot  \sum_{j\neq i}^{n-1} \vec u(2\phi_j) \\ \nonumber
&=&\frac{\vec V_2(p_{t})\cdot \vec V_2}{ n_{p_t} },
\eea
and normalization is obtained from
\bea
\vec Q_{2a} \cdot \vec Q_{2b} &=&  \sum_{i\in a\neq j\in b}^{n_{a}\sim n_{b}} \cos( 2[\phi_i -\phi_j]),
\eea
where $a,\, b$ denote two equivalent and disjoint partition elements (subevents) covering a detector acceptance ($n_a \simeq n_b \simeq n/2$). {If} the disjoint partition elements are perfectly correlated, and there are no {\em nonflow} contributions, then $\vec Q_{2a} \cdot \vec Q_{2b} =\vec V_{2a} \cdot \vec V_{2b} \simeq V_2^2 / 4$~\cite{flowmeth,gluequad} and
\bea \label{v22pt}
\frac{\langle \vec u(2\phi_{i\in p_t})\cdot  \vec Q_{2i}\rangle}{2\, \sqrt{\vec Q_{2a} \cdot \vec Q_{2b}}} &\simeq&  \frac{\vec V_2(p_{t})\cdot \vec V_2}{ n_{p_t} \, V_2} \equiv v_2\{2\}(p_t),
%\\ \nonumber
%&\equiv& v_2\{2\}(p_t),
\eea
explaining Eq.~(5) of~\cite{0409033}. If Eq.~(\ref{v22pt}) is multiplied top and bottom by $1/Q_2$ the EP method Eq.~(\ref{v2ep}) is approximated. The first relation in Eq.~(\ref{v22pt}) is approximate because of uncertain physical implications of the definition of (cut system for) partition elements $a$ and $b$. The definition of the $(a,b)$ partition may reduce nonflow contributions to $\vec Q_{2a} \cdot \vec Q_{2b} $ compared to $\vec V_2(p_t)\cdot \vec V_2$, leading to an undetermined systematic error in ratio $v_2(p_t)$.

This section demonstrates that $v_2\{EP\}(p_t) \approx v_2\{2\}(p_t)$ approximates a generic two-particle correlation analysis on azimuth, although motivating language and symbols (e.g., ``flow vector'' $\vec Q_2$) imply that the event-plane method {\em necessarily} relates to ``collective'' (hydrodynamic) phenomena. Multiplicities $n_{p_t}$ are elements of the histogrammed single-particle $p_t$ spectrum. The $p_t$ spectrum in the $v_2(p_t)$ denominator obscures interpretation and comparisons to theory, as shown in this analysis.

%%%%%%%%%
\section{${\bf v_2(p_t)}$ measurements} \label{v2meas}

$v_2$ data are described in the context of a thermalized, collectively-flowing bulk partonic medium probed by flow measurements and energetic scattered partons. Smaller-$p_t$ hadrons emerging from the bulk medium (possibly by coalescence of ``thermal'' partons) exhibit a pattern of flows. Larger-$p_t$ hadrons from parton fragmentation (possibly by coalescence of ``shower'' partons) reveal modification of fragmentation by the medium. Intermediate-$p_t$ hadrons may result from recombination of ``thermal'' and ``shower'' partons~\cite{quarkcoal,rudy,fries,greco}.

In the present analysis qualitative {\em conceptual} issues are of central importance. A simple and accurate data sample including both $p_t$ and mass dependence is used to demonstrate the algebraic structure of $v_2(p_t)$ and the basic properties of quadrupole spectra. Notable theory examples are included to explore the general relation of hydro theory to the quadrupole component in different manifestations. $v_2(p_t)$ data for pions, kaons and Lambdas are related to single-particle spectra for pions, kaons (interpolated) and protons. Proton and Lambda spectrum {\em shapes} are assumed equivalent for this analysis.

%\subsection{Conventional ${\bf v_2}$ description}

\begin{figure}[h]
  \includegraphics[width=1.65in,height=1.63in]{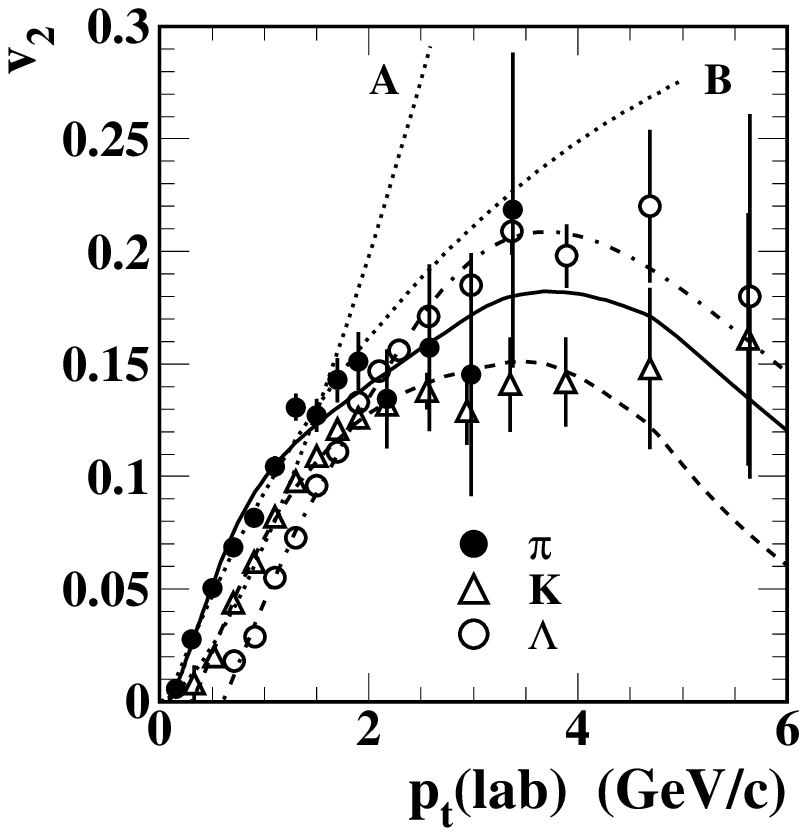}
   \includegraphics[width=1.65in,height=1.65in]{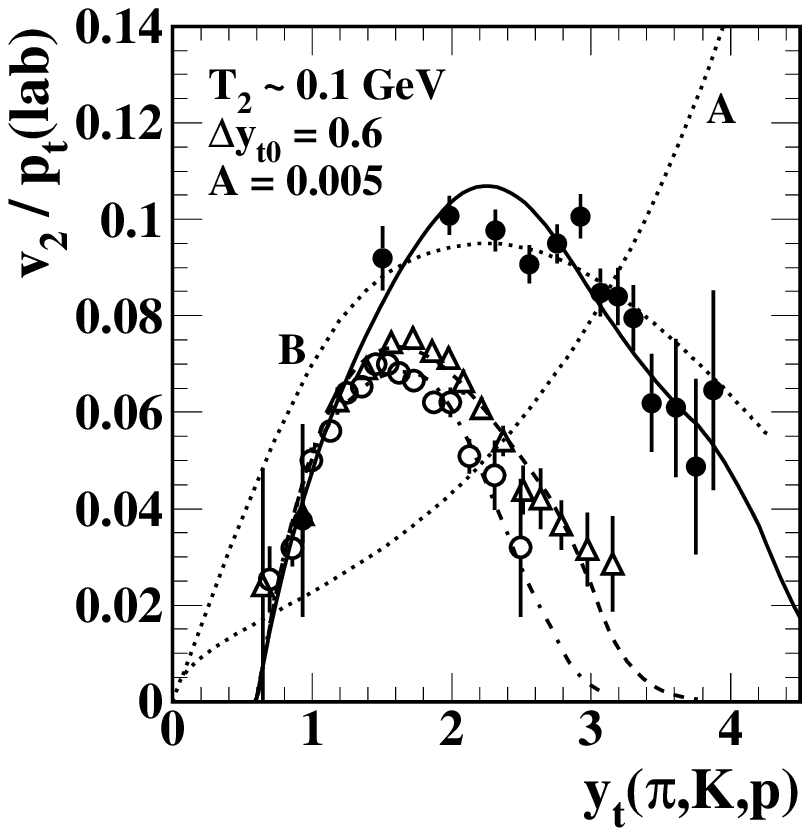}
\caption{\label{boost1a}
Left panel:  $v_2(p_t)$ data for three hadron species plotted in the usual format~\cite{v2pions,v2strange}.
 Right panel:   The same $v_2(p_t)$ data divided by $p_t$ in the lab frame suggest a universality on proper transverse rapidity for each hadron species---in particular the correspondence of data near $y_t = 1$. Dotted curves A and B in each panel are viscous hydro predictions from~\cite{teaney} and~\cite{rom} respectively. The three curves through data are derived from this analysis.
} 
%boost1a, boost1bb
 \end{figure}

Fig.~\ref{boost1a} (left panel) shows data from $v_2(p_t)$ analysis of identified mesons (pions, kaons) and baryons (Lambdas) from minimum-bias 200 GeV Au-Au collisions~\cite{v2pions,v2strange}. The mass trend at smaller $p_t$ (massive hadrons have smaller $v_2$) is commonly interpreted to imply collective flow (hydrodynamics). At larger $p_t$ $v_2$ data are said to ``saturate,'' following a nearly constant trend beyond 4 GeV/c~\cite{v2sat}. 

Hydrodynamic models provide a semi-quantitative description at smaller $p_t$ but fail at larger $p_t$ (hydro models overpredict $v_2$ at larger $p_t$)~\cite{v2over}. 
The dotted curves in each panel are viscous hydro predictions with zero-viscosity limit for pion $v_2(p_t)$ (A from~\cite{teaney}, B from~\cite{rom}). The systematics of viscous hydro predictions compared to data are interpreted as evidence for a fluid medium with very small viscosity (``perfect liquid'')~\cite{teaney,hirano2,csernai}. 

%%%%%%%%%%
%%%%%%%%%%

%\subsection{Alternative format and description}

Fig.~\ref{boost1a} (right panel) shows the same data in the form $v_2(p_t) / p_t(\text{lab frame})$ plotted {\em vs} $y_t(\pi,K,p)$ (proper rapidity for each hadron species), where transverse rapidity $y_t \equiv \log\{(m_t + p_t) / m_0\}$. The first Lambda point at $y_t \sim 0.57$ (not visible) is slightly negative but consistent with zero. The simple transformation, revealing peaked distributions with similar amplitudes and {\em common left edges} for the three hadron species, suggests that more information can be extracted from existing $v_2$ data with a generalized analysis method. 

Data distributions in the right panel taken together imply that the three hadron species are emitted from a common moving (boosted) source, as demonstrated below. The three curves from this analysis passing through data in each panel are based on that hypothesis. The relevant model parameters are summarized in the panel. The dotted hydro curves in the right panel~\cite{teaney,rom} deviate significantly from the pion data trend. However, the same curves compared to pion data in the left panel have been cited to imply small medium viscosity and formation of a ``perfect liquid'' at RHIC. The relation of the hydro curves to data is discussed in Sec.~\ref{theorycomp}.

%%%%%%%%%
\section{Single-particle spectra} \label{singlespec}

The first step of this analysis is to obtain the single-particle azimuth-averaged 3D spectrum $\rho_0 = Q_0(y_t)/2\pi$ (denominator of $v_2[p_t]$) for three hadron species from Au-Au collisions at 200 GeV which provides a context for the quadrupole component (numerator of $v_2[p_t]$). Single-particle spectra may include a quadrupole contribution  which can be used to estimate the absolute quadrupole yield. Two-component spectrum analyses of p-p and Au-Au spectra reported in~\cite{ppprd,2comp} are used to construct minimum-bias spectra compatible with the data in Fig.~\ref{boost1a}. 

\subsection{Spectrum notation}

The 3D single-hadron density on momentum averaged over one unit of pseudorapidity $\eta$ about mid-rapidity is
\bea \label{densities}
\rho_0(x_t,\phi) &\equiv& \frac{1}{ x_t}\frac{d^3n}{dx_t\, d\eta\,d\phi} \\ \nonumber
\rho_0(x_t) &\equiv& \frac{1}{2\pi x_t}\frac{d^2n}{dx_t\, d\eta},
\eea
where the second line is averaged over azimuth. Transverse measure $x_t$ is $p_t,\, m_t$ or $y_t$. Transformations between densities require Jacobians $dy_t/ dp_t = 1/m_t$ and $dy_t/ dm_t = 1/p_t$. For reference, $\rho_{NN} = d^2n_{NN}/d\eta d\phi \sim 2.5/2\pi$ is the $\eta$- and $\phi$-averaged, $p_t$-integrated 2D hadron density at mid-rapidity for 200 GeV NSD N-N collisions. 

It is sometimes useful to plot all hadron species on pion rapidity denoted by $y_{t\pi} \approx \ln(2p_t / m_\pi)$ or (for plot axes) $y_t(\pi)$. $y_{t\pi}$ is then simply a logarithmic measure of $p_t$ providing better visual access to spectrum structure. When relativistic transformations (boosts) are important the proper $y_t$ for each hadron species should be used, denoted by variable $y_t$ with no qualification and plot axis labels $y_t(\pi,K,p)$.

Comparison of results on transverse variables $p_t$, $m_t$ and $y_t$, as in this study, is essential to distinguish different physical mechanisms. For thermal spectra $m_t$ is preferred. For boosted systems proper $y_t$ for each hadron species is preferred. For parton fragmentation (minijets) $p_t$ would reflect the common underlying parton spectrum, but $y_{t\pi}$ ($\sim \ln(2p_t / m_\pi)$) provides better visual access to structure. Analysis of spectra on a single plotting variable may confuse several dynamical mechanisms.

\subsection{Glauber model and multiplicity definitions}

The Glauber model of A-A collisions defines several A-A geometry parameters~\cite{centmeth}. For A-A impact parameter $b$, $n_{part}/2$ is the corresponding average number of participant nucleon pairs and $n_{bin}$ is the average number of N-N binary collisions (for a specified scattering process). Some hadron production processes are proportional to $n_{part} / 2$ (soft), and some are proportional to $n_{bin}$ (hard). The combination comprises the two-component model of hadron production, which describes N-N collisions well~\cite{ppprd} and serves as a reference in A-A collisions~\cite{2comp}. $\nu \equiv 2\, n_{bin} / n_{part}$, the mean participant pathlength in number of encountered nucleons, is a geometry parameter used to measure A-A centrality.

$n_{ch}$ is the total charged-particle multiplicity in one unit of $\eta$ at mid-rapidity. The total multiplicity associated with the quadrupole component is $n_{ch2}$. The quadrupole multiplicity associated with hadron species X is $n_{chX2}$. Quadrupole multiplicities should not be confused with quadrupole L\'evy distribution shape parameter $n_2$ or $n_{X2}$.

Ambiguities in the normalizations of measured spectra are noted in~\cite{2comp}, specifically the centrality dependence of integrated $n_{ch}$ compared between experiments. The present analysis concerns minimum-bias $v_2(p_t)$ data for which an average over centrality is implicit. The associated normalization uncertainty in the averaged single-particle spectra is about 20\%. However, normalization uncertainty is not relevant to the present analysis which refers only to relative spectrum shapes.

\subsection{Two-component spectrum model} \label{specmodels}

The two-component (soft+hard) model of hadron spectra provides a compact and accurate description of p-p and Au-Au collisions~\cite{ppprd,2comp}. The soft component is interpreted as longitudinal participant-nucleon fragmentation. The hard component at mid-rapidity is interpreted as {\em minimum-bias} large-angle scattered parton fragmentation (minijets), which can also be described as hadrons emitted from a radially-boosted source. The open question for any observed boost phenomenon is what {\em physical mechanism} produced the boost.

The two-component models of pion, kaon and proton spectra (3D densities {\em per participant pair}) at 200 GeV are summarized by
\bea  \label{summ}
\frac{2}{n_{part}} \rho_{0\pi} \hspace{-.04in} &=& \hspace{-.04in} \frac{0.85\, \rho_\text{NN}}{1.012} \{ S_{0\pi}  \hspace{-.02in}+ \hspace{-.02in} 0.012\, \nu\, r_{AA\pi}\,  H_{0\pi} \} \\ \nonumber
\frac{2}{n_{part}} \rho_{0K} \hspace{-.04in} &=& \hspace{-.04in} \frac{0.09\, \rho_\text{NN}}{1.16} \{ S_{0K}  \hspace{-.02in}+ \hspace{-.02in} 0.16\, \nu\, r_{AAK}\,  H_{0K} \} \\ \nonumber
\frac{2}{n_{part}} \rho_{0p} \hspace{-.04in} &=& \hspace{-.04in} \frac{0.06\, \rho_\text{NN}}{1.12} \{ S_{0p} \hspace{-.02in} + \hspace{-.02in} 0.12\, \nu\, r_{AAp}\,  H_{0p} \},
\eea
with the differential form of $\rho_{0X}(y_{t})$ defined in Eq.~(\ref{densities}). Unit-integral model functions  $S_{0X}(y_{t})$ and $H_{0X}(y_{t})$ and hard-component ratios $r_{AAX}(y_t;\nu)$ for pions and protons are defined in~\cite{2comp}. The $r_{AAX}$ represent all deviations from the N-N + Glauber two-component {\em linear reference}. Kaon model functions were estimated by interpolation for this analysis.

%%%%%%%%%%
 \begin{figure}[h]
 \includegraphics[width=1.65in,height=1.63in]{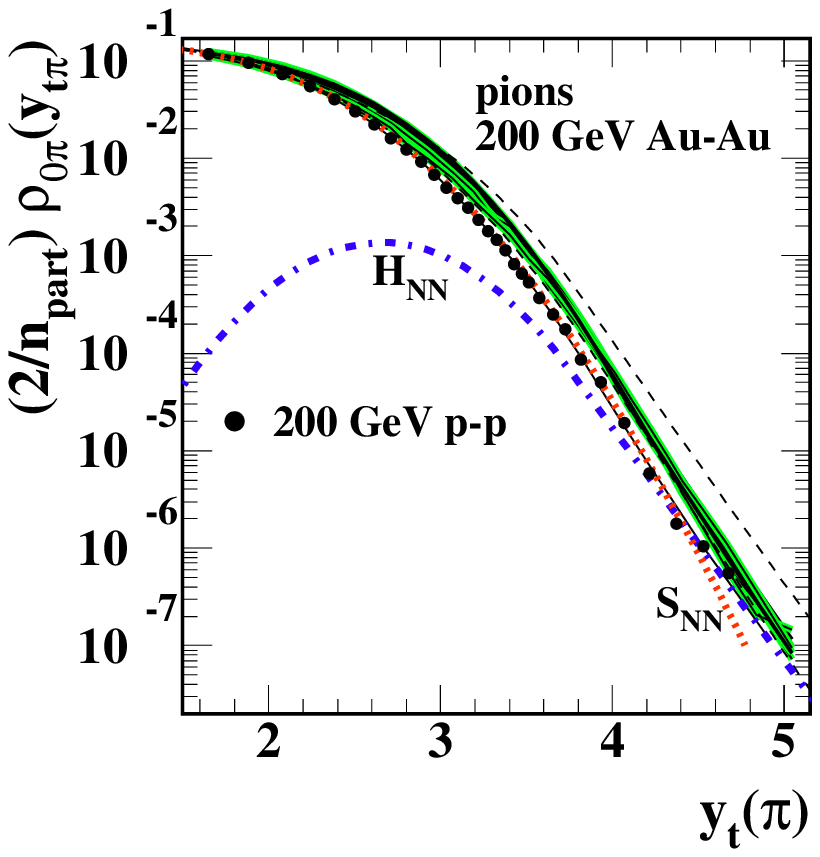}
   \includegraphics[width=1.65in,height=1.65in]{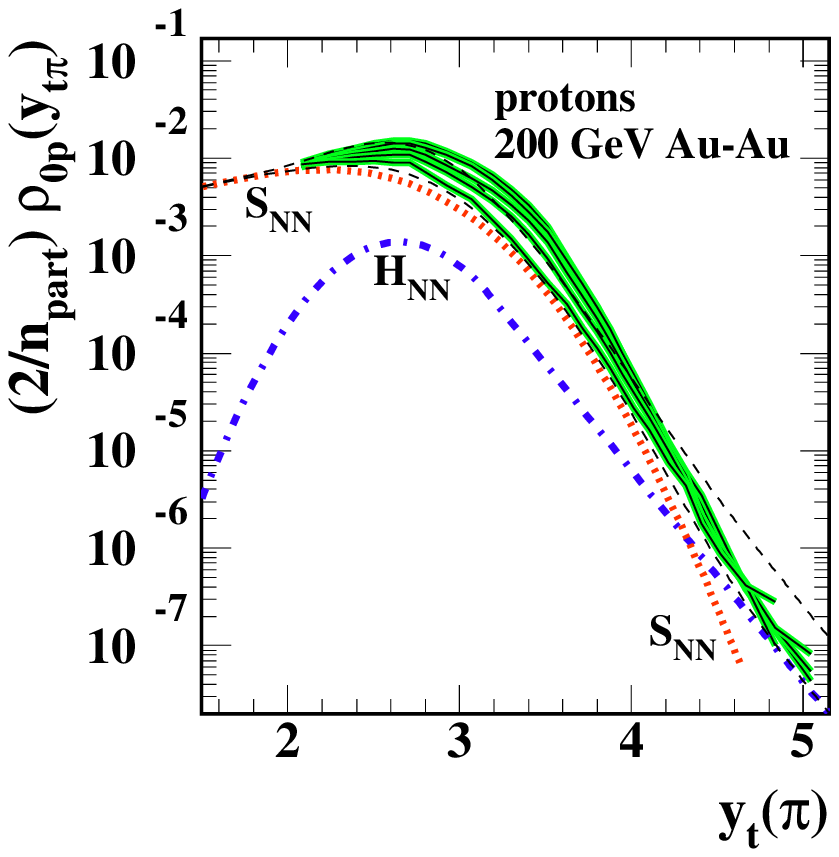}
\caption{\label{aaspec10}
(Color online) Summary of pion (left) and proton (right) per-participant-pair single-particle spectra from Au-Au collisions at 200 GeV and five centralities~\cite{2comp}. $H_{NN}$ is the hard component (minimum-bias transverse parton fragmentation) and $S_{NN}$ is the soft component (longitudinal nucleon fragmentation), both inferred for N-N collisions. The solid points in the left panel represent the NSD p-p spectrum~\cite{ppprd}.
 } 
%boost10bx, boost11bx (from aaspectra)
 \end{figure}
%%%%%%%%%%

\begin{figure}[h]
 \includegraphics[width=3.3in,height=1.63in]{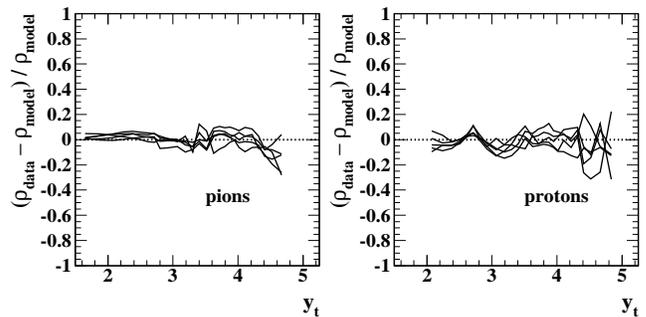}
\caption{\label{aaspec25}
Relative residuals (data $-$ model) / model for pions (left) and protons (right) from five centralities of Au-Au collisions at  $\sqrt{s_{NN}} = 200$ GeV. The two-component spectrum model with modification factors $r_{AA}$ inferred in~\cite{2comp} describes data to the statistical limits from 0.5 to 10 GeV/c ($y_t \in [2,5]$).
 } 
%boost10bx, boost11bx (from aaspectra)
 \end{figure}

Model spectra describing pion and proton data are summarized in Fig.~\ref{aaspec10}. Reference soft components $S_0$ (unit-normal distributions not shown) are L\'evy distributions on $m_t$ transformed to {\em pion} $y_t$. The transformation strategy is discussed in~\cite{2comp}. Reference hard components $H_0$ (also unit-normal distributions not shown) are Gaussians on $y_t$ with exponential tails $\propto \exp(-n_{y_t}\, y_t)$ representing expected QCD power law $p_t^{-n_h}$ required by data above $p_t \sim 6$ GeV/c ($y_{t\pi} \sim 4.5$). Distributions $S_{NN}$ and $H_{NN}$ have the same forms but integrate to hard- and soft-component hadron numbers $n_s$ and $n_h$ with $n_s + n_h = n_{ch}$ for N-N collisions~\cite{ppprd}. By hypothesis, the soft component for Au-Au collisions remains fixed at the N-N reference. Deviations of hard component $H_{AA}$ from its N-N reference are measured by ratio $r_{AA} = H_{AA} / H_{NN}$. The model functions describe the shapes of the data spectra at the few-percent level over the $y_t$ interval relevant to this analysis. The quality of the description is indicated by the relative residuals in Fig.~\ref{aaspec25}.

%%%%%%%%%%
%%%%%%%%%%

\subsection{The quadrupole spectrum component}

Interpretation of the azimuth quadrupole spectrum raises a significant question: Does the azimuth quadrupole ($v_2$ data) represent {\em modulation} of a spectrum component  existing in N-N collisions (e.g., the soft component), or does a new radially-boosted {\em net} source of hadrons modulated on azimuth emerge in A-A collisions? Does that component extrapolate back to N-N collisions?

The $\eta$-averaged three-component 3D spectrum on $(x_t,\phi)$ for $x_t = m_t$ or $y_t$ can be expressed as
\bea
\rho(m_t,\phi) &= & \rho_0(m_t;T_0) + \rho_2(m_t;T_2,\beta_{t}[\phi]) \\ \nonumber
\rho(y_t,\phi) &= & \rho_0(y_t;\mu_0) + \rho_2(y_t;\mu_2,\Delta y_{t}[\phi]),
\eea
where $\rho_2$ is a possible quadrupole (third) component from a radially-boosted source. Parameters $\beta_{t}(\phi)$ and $\Delta y_{t}(\phi)$ represent a conjectured azimuth-dependent radial boost of the third component. The first term $\rho_0(y_t;\mu_0)$ is the two-component spectrum from~\cite{2comp}. Quadrupole term $\rho_2$ may represent a new particle source, a modification of the N-N soft component, of the hard component, or an interaction between them. To clarify we must estimate the shape and absolute magnitude of the quadrupole spectrum component from $v_2(p_t)$ data and compare them with measured azimuth-averaged $y_t$ spectra.

%%%%%%%%%
\section{Boosted hadron sources} \label{boost}

The second step of this analysis is to define the kinematics of nearly-thermal hadron spectra from moving (boosted) sources, essentially the blast-wave model~\cite{blastwave,starblast} related to the Cooper-Frye description of moving (expanding) particle sources~\cite{cooperfrye}. I consider only monopole and azimuth-quadrupole $p_t$ and $y_t$ spectrum components. For simplicity ``thermal'' spectra are described in the boosted frame by Maxwell-Boltzmann exponentials on $m_t$. The description can be generalized to L\'evy distributions on $m_t$ for accurate descriptions of data. The intent is to provide a general description of hadron production from a source including (but not restricted to) a radially-boosted component with azimuth variation. 

\subsection{Radial boost kinematics}

The four-momentum components of a boosted source are first related to transverse rapidity $y_t$. The boost distribution is assumed to be a single value for simplicity.  The {\em particle} four-momentum components are $m_t = m_0 \cosh(y_t)$ and $p_t = m_0\sinh(y_t)$. The {\em source} four-velocity (boost) components are $\gamma_t = \cosh(\Delta y_t)$ and $\gamma_t \, \beta_t= \sinh(\Delta y_t)$, with $\beta_t= \tanh(\Delta y_t)$. Boost-frame variables are defined in terms of lab-frame variables by
\bea \label{boostkine}
m'_t  &\equiv& m_0\, \cosh(y_t - \Delta y_{t}) =\gamma_{t}\, (m_t - \beta_{t}\, p_t)  \\ \nonumber
 &=& m_t\, \gamma_{t} \{ 1 - \tanh(y_t)\, \tanh(\Delta y_{t})  \} \\ \nonumber
p'_t &\equiv&   m_0\, \sinh(y_t - \Delta y_{t}) =\gamma_{t}\, (p_t - \beta_{t}\, m_t) \\ \nonumber
&=& m_t\, \gamma_{t} \{  \tanh(y_t) - \tanh(\Delta y_{t})  \} 
%\frac{\tilde p_t}{m_t}   &=& \frac{\tanh(y_t) - \tanh(\Delta y_{t})}{1 - \tanh(\Delta y_{t})} \\ \nonumber
\eea
with $p'_t$ denoted $p_t$(boost) in figures. 
%and $p'_t / m'_t = \tanh(y_t - \Delta y_{t})$.
  
\begin{figure}[h]
   \includegraphics[width=1.65in,height=1.65in]{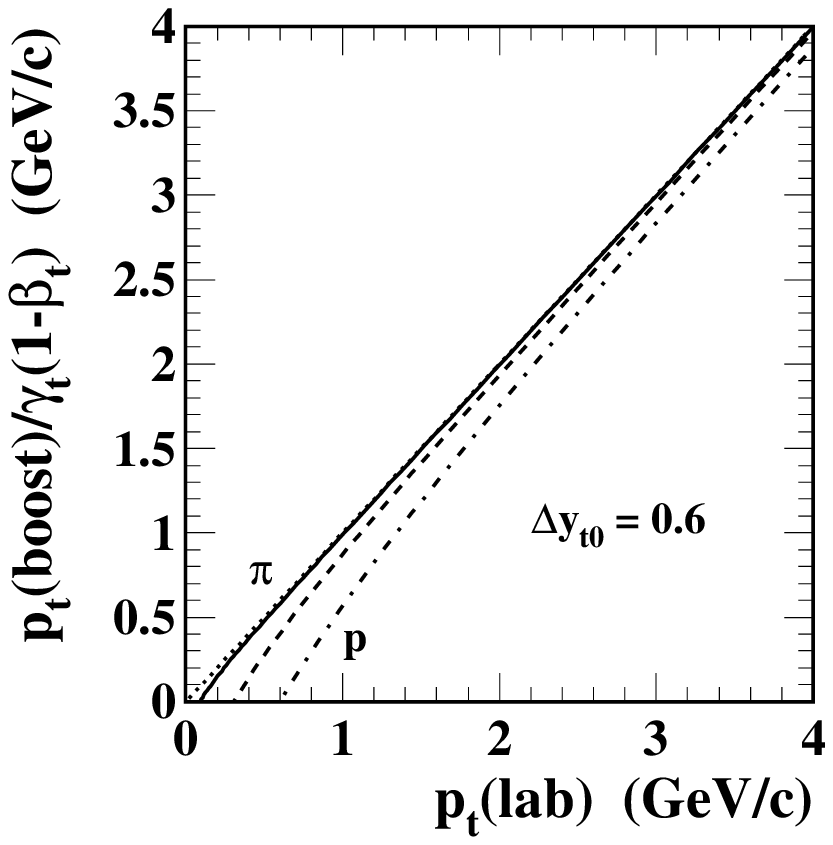}
  \includegraphics[width=1.65in,height=1.63in]{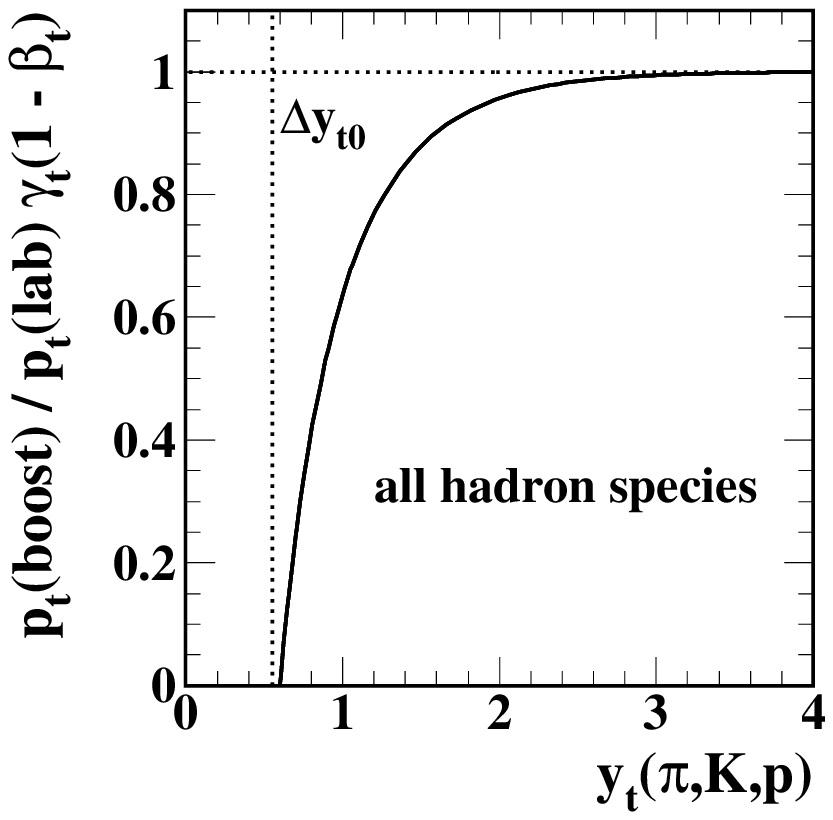}
\caption{\label{boost3}
Left panel: $p'_t$ ($p_t$ in the boost frame) {\em vs} $p_t$ in the lab frame. Factor $\gamma_{t}(1 - \beta_{t})$ in the denominator insures that the combination $\rightarrow p_t$ for large $p_t$.
Right panel:  The same ratio is plotted on proper $y_t$ for each hadron species, demonstrating a fundamental relationship applicable to any hadron species.
 } 
%boost3cc, boost3d
 \end{figure}

Fig.~\ref{boost3} (left panel) relates $p'_t \rightarrow p_t(\text{boost})$ to $p_t \rightarrow p_t(\text{lab})$. The main source of the mass trend of $v_2(p_t)$ at small $p_t$, interpreted as ``hydro'' behavior, is a simple kinematic effect as seen at lower left. The mass systematics hold for any boosted nearly-thermal hadron source independent of boost mechanism (i.e., hydrodynamics is not required). The {\em intercepts} ($p'_t = 0$) of the three curves, given by $p_{t0} = m_0 \sinh(\Delta y_{t})$, are important for discussion of the hydro interpretation of $v_2(p_t)$. 

%%%%%%%%%%
%%%%%%%%%%

Fig.~\ref{boost3} (right panel) relates $p'_t$ to transverse rapidity $y_t(\pi,K,p)$ and illustrates one reason why plots on $y_t$ are a major improvement over $p_t$ or $m_t$. Normalized $p'_t / p_t$  
\bea
 \frac{p'_t}{p_t\, \gamma_{t}(1 - \beta_{t})} 
    &=& \frac{1 - \beta_t /\tanh(y_t) }{1 - \beta_t}
\eea
increases from zero at monopole boost $\Delta y_{t0} $ and follows a {\em universal curve on $y_t$} to unit value for any hadron species. Thus, normalized $p'_t$ goes asymptotically to $p_t$ for large $p_t$ (or $y_t$) independent of boost. The form in Fig.~\ref{boost3} (right panel) is important for interpreting $v_2(p_t)$ data as {\em spectra}. In the present study we find $\Delta y_{t0} \sim 0.6 \sim \gamma_t\, (1 - \beta_t)$, common to three hadron species.

\subsection{Radially-boosted thermal spectra}

The simple blast-wave model invoked here assumes longitudinal-boost-invariant normal emission from an expanding cylinder, eliminating the need for Bessel functions $K_1$ and $I_0$ arising from integrals over $y_z$ and $\phi$~\cite{schned}. Slope parameter $T$ for $m_t$ spectra and thermal parameter $\mu = m_0 / T$ for $y_t$ spectra are defined.  Boosted spectra on $y_t$ and $m_t$~\cite{cooperfrye} are
\bea
\rho(y_t;\mu,\Delta y_t)  \hspace{-.05in} &=&  \hspace{-.05in} A_{y_t}\,\exp\{ -\mu\, [ \cosh(y_t - \Delta y_t) - 1]\} \\ \nonumber
\rho(m_t;T,\beta_t) \hspace{-.05in} &=&  \hspace{-.05in} A_{m_t}\, \exp\{- [\gamma_t\, (m_t - \beta_t\, p_t) - m]/T \},
\eea
providing a simplified description of ``thermal'' radiation from a radially-boosted cylindrical  source. Applications require a specific radial boost model $\Delta y_t(r,\phi)$.

\subsection{Radial boost models} \label{boostc}

In a nuclear collision there are (at least) two possibilities for the radial boost model: 1) a monolithic, thermalized, collectively-flowing hadron source (``bulk medium'') with complex transverse flow distribution dominated by monopole (radial flow or Hubble expansion) and quadrupole (elliptic flow) azimuth components~\cite{starblast}; and 2) multiple hadron sources, some with azimuth-modulated transverse boost. Hadrons may emerge from a radially-fixed source (soft component), from parton fragmentation (hard component), and possibly from a source with radial boost varying smoothly on azimuth, including monopole and quadrupole components. Case 2 is assumed for this analysis, but both possibilities are  reconsidered in light of analysis results. 

A radial boost with monopole and quadrupole components is described by
\bea \label{quadboost}
\Delta y_{t}(\phi) &=& \Delta y_{t0} + \Delta y_{t2}\, \cos(2\Delta \phi_r ) \\ \nonumber
\beta_t(\phi) &=&  \tanh(\Delta y_t[\phi]) \\ \nonumber
&\simeq&\beta_{t0} + \beta_{t2}\, \cos(2\Delta \phi_r ),
\eea
with $\Delta y_{t2} \leq \Delta y_{t0}$ for {\em positive-definite boost}. The convention $\Delta \phi_r \equiv \phi - \Psi_r$ is adopted for more compact notation. Monopole boost component $\Delta y_{t0}$ is easy to extract from data, but quadrupole component $\Delta y_{t2}$ is less accessible. $\Delta y_{t0}$ could be interpreted as a ``radial flow'' but may apply to only a small fraction of produced hadrons. The quadrupole boost magnitude should reflect the eccentricity $\epsilon$ of the A-A collision geometry.
%, so $\Delta y_{t2} \equiv \xi_2\, \epsilon$, defining $\xi_2$.

%%%%%%%%
\section{Azimuth Quadrupole Component} \label{quadcomp}

The third step of this analysis is to relate the azimuth quadrupole spectrum component algebraically to experimental $v_2$ data.  I assume that  1) the quadrupole component arises from a hadron source with azimuth-dependent radial boost distribution $\Delta y_t(\phi)$, 2) the quadrupole source {\em may} produce only a small fraction of the hadrons in a collision, and 3) the quadrupole spectrum may {appear} to be thermal in the boost frame and {\em may be} independent of the soft and hard spectrum components. 

\subsection{Quadrupole-component model}

Given those assumptions the $\eta$-averaged 3D spectrum at midrapidity for hadrons associated with the quadrupole component is modeled by
\bea \label{fullquad}
\rho_2(y_t,\phi) \hspace{-.07in} &=& \hspace{-.07in} A_{2y_t} \exp\{ -\mu_2[\cosh(y_t - \Delta y_{t}(\phi)) - 1]  \} \\ \nonumber
\rho_2(m_t,\phi) \hspace{-.07in} &=& \hspace{-.07in} A_{2m_t} \exp\{ -\left(\gamma_t(\phi)[m_t - \beta_t(\phi) p_{t} ] - m_0\right)  /T_2\},
\eea
where a Maxwell-Boltzmann (M-B) distribution for a locally-thermal source is assumed for simplicity, and $\mu_2 = m_0 / T_2$ for the quadrupole. The procedure below may be applied to a more general function such as a L\'evy distribution.  $\Delta y_{t}(\phi)$ defined in Eq.~(\ref{quadboost}) represents fixed monopole and quadrupole boost components.

%%%%%%%%%%
 \begin{figure}[h]
  \includegraphics[width=1.65in,height=1.63in]{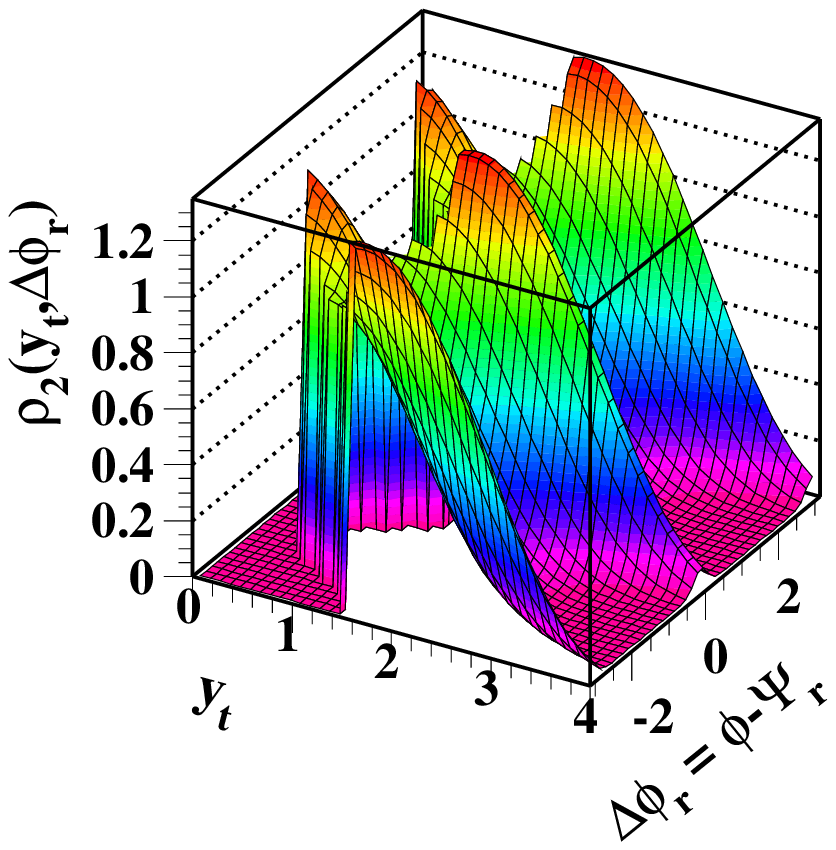}
   \includegraphics[width=1.65in,height=1.65in]{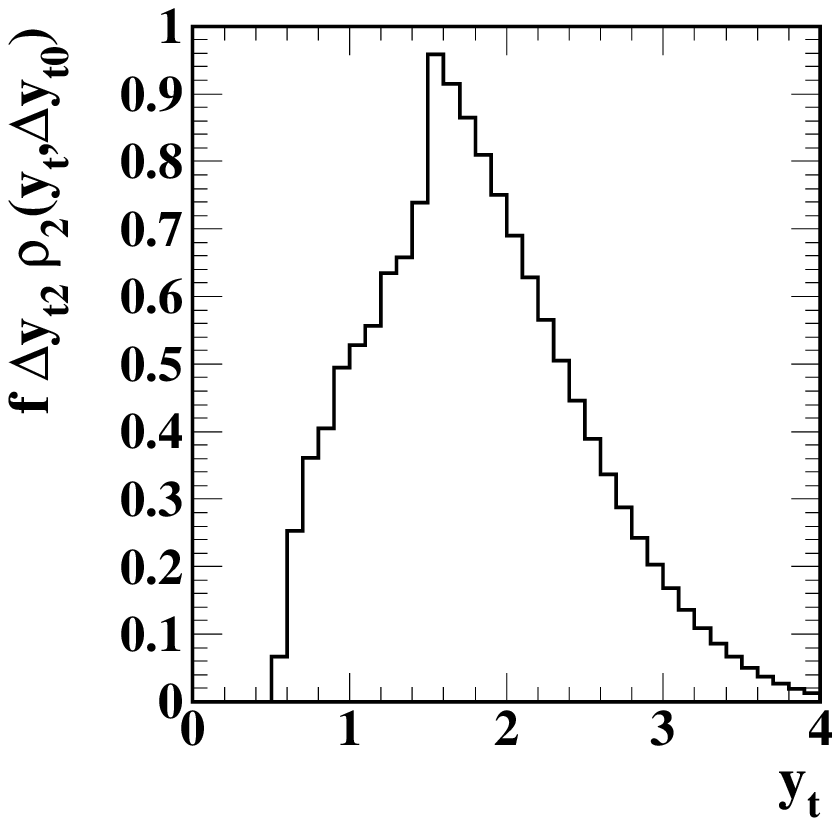}
\caption{\label{boost14a}
Left panel:   (Color online) The quadrupole component modeled by a thermal spectrum boosted by $\Delta y_t(\phi)$ containing monopole and quadrupole terms: monopole boost $\Delta y_{t0} = 1$ and quadrupole boost amplitude $\Delta y_{t2} = 0.5$. Right panel: projection of the left panel onto $y_t$ revealing the edge structure.
 } 
%boost14a-5, boost14b-5
 \end{figure}
%%%%%%%%%%

Fig.~\ref{boost14a} (left panel) illustrates the form of $\rho_2(y_t,\phi)$ relative to reference or reaction-plane angle $\Psi_r$, with $\Delta y_{t0} = 1$ and  $\Delta y_{t2} = 0.5$. Fig.~\ref{boost14a} (right panel) shows the projection of $\rho_2(y_t,\phi)$ onto $y_t$. The half-maximum point of  the left edge of the quadrupole spectrum is at monopole boost $\Delta y_{t0}$. The projection suggests that the shape of the left edge might reveal quadrupole boost amplitude $\Delta y_{t2}$ if resolved accurately. However, the small-$y_t$ region is experimentally difficult. The left edge is also affected by variations of $\Delta y_{t0}$  within a centrality bin. Accurate data would be needed to determine edge details.

\subsection{Quadrupole Fourier amplitude} \label{negproton}

\begin{figure}[h]
\includegraphics[width=3.3in,height=1.63in]{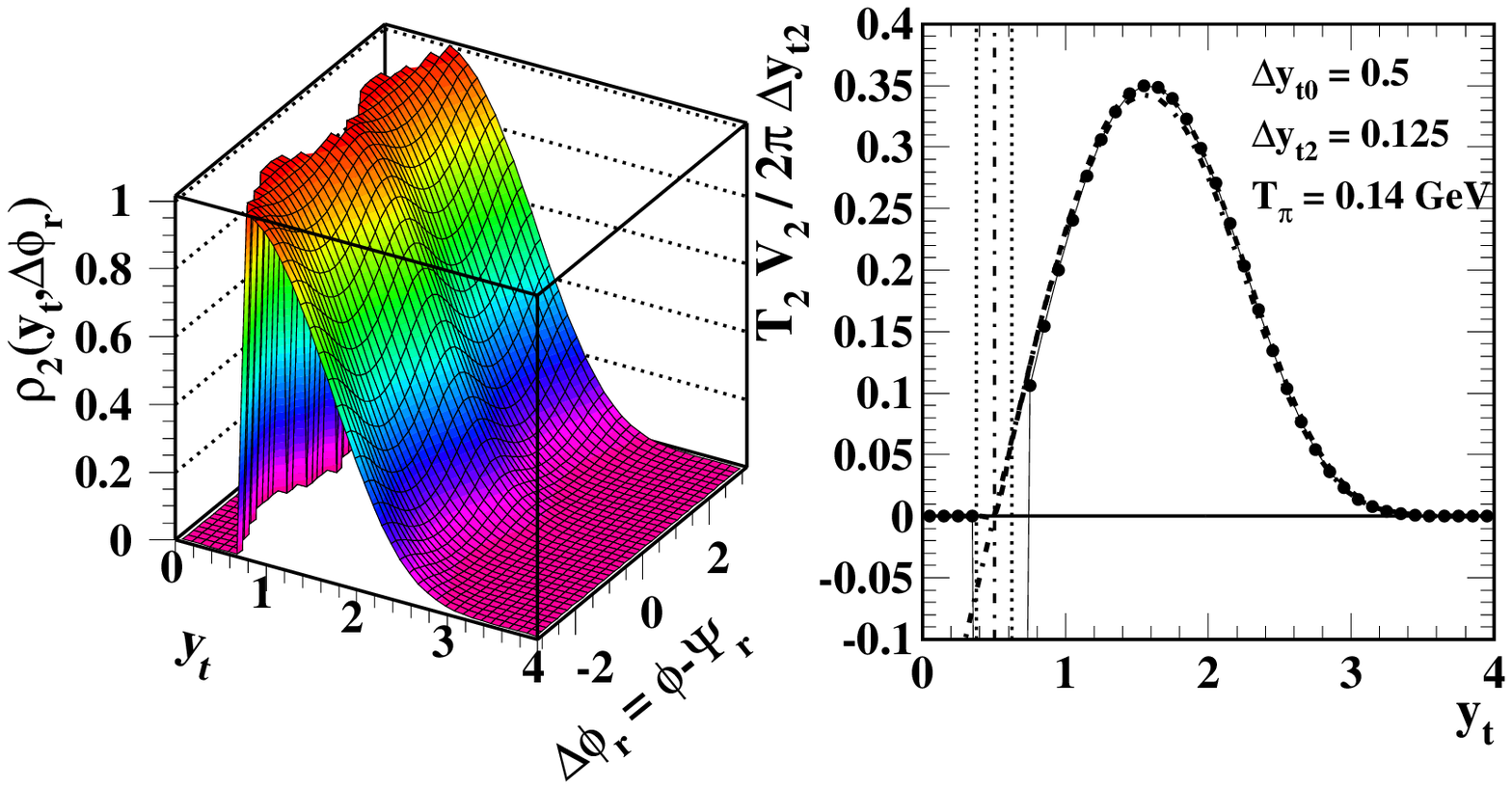}
   \includegraphics[width=3.3in,height=1.65in]{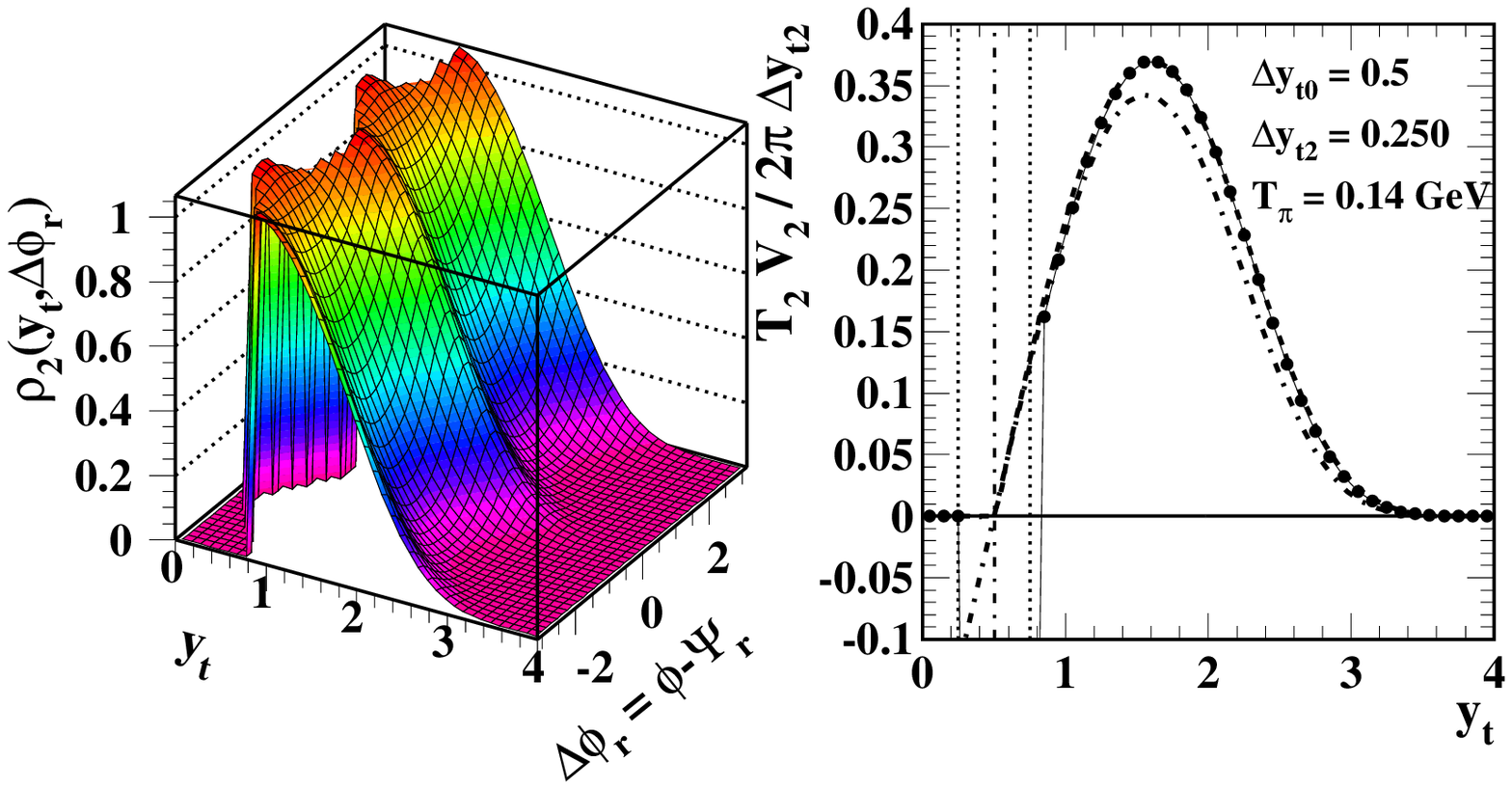}
\caption{\label{boost14ab}
Left panels: (Color online) The quadrupole component modeled as in Fig.~\ref{boost14a} for $\Delta y_{t0} = 0.5$ and two values of quadrupole boost amplitude $\Delta y_{t2}$.
Right panels:   Corresponding Fourier amplitudes $V_2(y_t)$ from Eq.~(\ref{v2int}) normalized by quadrupole boost amplitudes $\Delta y_{t2}$. Widths of the negative regions are $2 \Delta y_{t2}$. The dash-dot curve is the quadrupole spectrum $\rho_2(y_t;\Delta y_{t0})$.
 } 
%boost14ab-125, boost14ab-250
 \end{figure}

Fig.~\ref{boost14ab} (left panels) shows unit-amplitude $\rho_2(y_t,\phi)$ for $\Delta y_{t0} = 0.5$ and quadrupole boost amplitude $\Delta y_{t2} = 0.125,\, 0.250$. We can now obtain the relation between inferred quadrupole Fourier amplitude $V_2(y_t)$ and quadrupole spectrum $\rho_2(y_t;\Delta y_{t0})$. The Fourier amplitude is defined by
\bea \label{v2int}
V_2(y_t;\Delta y_{t0},\Delta y_{t2}) \equiv  \int_{-\pi}^{\pi} d\phi\, \rho_2(y_t,\phi)\,\cos(2 \Delta \phi_r)
\eea
assuming that reaction-plane angle $\Psi_r$ is known and $\rho_2(y_t,\phi)$ represents all $m = 2$ azimuth dependence in the single-particle spectrum. 

%%%%%%%%%%
%%%%%%%%%%

Fig.~\ref{boost14ab} (right panels) shows the integral in Eq.~(\ref{v2int}) times $T_2 / 2\pi \Delta y_{t2}$ as a histogram (points and thin solid curve). The Fourier amplitudes peak at $y_t \sim 1.5$ and fall toward zero at $y_t = \Delta y_{t0}$. Similar amplitudes at the peak confirm that the integral is $\propto \Delta y_{t2}$. A negative undershoot centered at $\Delta y_{t0}$ (vertical dash-dot line) with width $\sim 2\Delta y_{t2}$ appears because the phase of the sinusoid changes by $\pi$ in traversing from one side of the mode of $\rho_2$ on $y_t$ ($y_t \sim \Delta y_{t0} + \Delta y_{t2}$) to the other. Negative values of $v_2$ do not {\em require} collective flow of a medium, only boost of the quadrupole component. The other curves are described below.

\subsection{Factoring $V_2(y_t)$}

Eq.~(\ref{v2int}) can be factored to isolate the underlying quadrupole spectrum $\rho_2(y_t;\Delta y_{t0})$, the  subject of this paper. Invoking the $\Delta y_{t}(\phi)$ model defined above and referring back to Eq.~(\ref{boostkine}) I expand the $\cosh$ term in the boosted M-B distribution of Eq.~(\ref{fullquad}) as
\bea
\cosh(y_t - \Delta y_t[\phi]) - 1  &=&  \cosh(y_t - \Delta y_{t0}) - 1 + \\ \nonumber
 && \hspace{-1.05in} \cosh(y_t - \Delta y_{t0}) \{\cosh(\Delta y_{t2}\, \cos[2\Delta \phi_r]) -1 \}+ \\ \nonumber
&& \hspace{-1.05in} \sinh(y_t - \Delta y_{t0}) \sinh(\Delta y_{t2}\, \cos[2\Delta \phi_r]) 
\eea
The three terms correspond to three factors of $\rho_2(y_t,\phi)$.
\bea \label{factor}
\rho_2(y_t,\phi) \hspace{-.0in} &=& A_2\, \exp\{ -(m'_t - m_0) / T_2 \}\times \\ \nonumber
&& \hspace{-.2in} \exp\{m'_t \, [\cosh( \Delta y_{t2}\, \cos[2 \Delta \phi_r]) - 1]/T_2\} \times \\ \nonumber
&& \hspace{-.2in} \exp\{p'_t \, \sinh( \Delta y_{t2}\, \cos[2 \Delta \phi_r]) /T_2\} \\ \nonumber
&\equiv& \rho_2(y_t;\Delta y_{t0})\times F_1(y_t,\phi;\Delta y_{t0} ,\Delta y_{t2})\times \\ \nonumber
&&F_2(y_t,\phi;\Delta y_{t0} ,\Delta y_{t2}).
\eea  
The last line defines azimuth-dependent factors $F_1(y_t,\phi)$ and $F_2(y_t,\phi)$ in terms of monopole and quadrupole components of the radial boost. The objective is quadrupole spectrum component $\rho_2(y_t;\Delta y_{t0})$ emitted from the boosted particle source as one factor of measured Fourier amplitude $V_2(p_t)$ inferred from $v_2(p_t)$ data. 

\begin{figure}[h]
  \includegraphics[width=1.65in,height=1.63in]{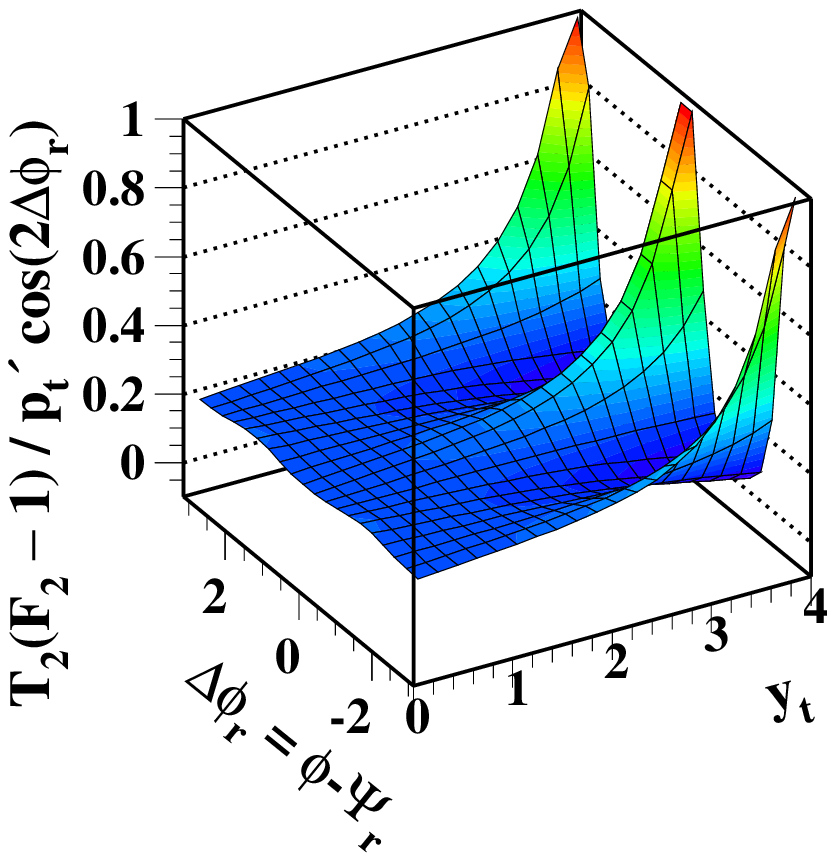}
  \includegraphics[width=1.65in,height=1.63in]{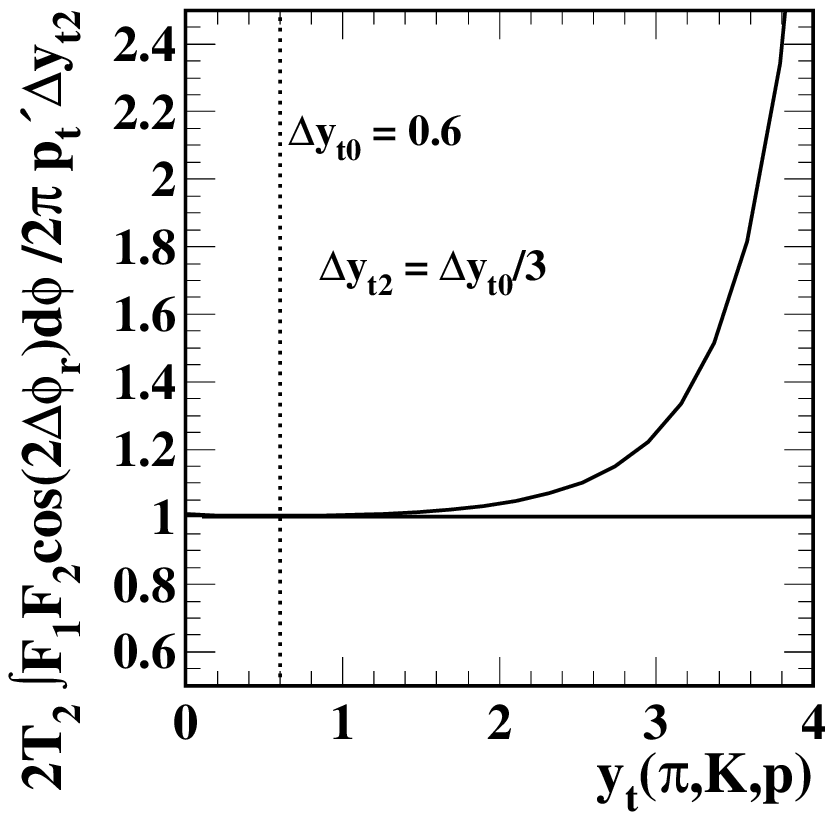}
\caption{\label{boost13a}
Left panel:   (Color online) The structure of factor $F_2$ in Eq.~(\ref{factor}), dominated by $\Delta y_{t2}$ at smaller $y_t$. 
Right panel: The structure of $O(1)$ factor $f(y_t;\Delta y_{t0},\Delta y_{t2})$ defined in Eq.~(\ref{fint}).
 } 
%boost13a, boost13b
 \end{figure}

The quadrupole boost dependence is contained in integral $\frac{1}{2\pi}\int_{-\pi}^{\pi}d\phi\, F_1(y_t,\phi)\, F_2(y_t,\phi)\, \cos(2 \Delta \phi_r)$. The leading azimuth dependence of $F_1 - 1$ is $\cos^2(2 \Delta \phi_r)$ which does not contribute appreciably to the integral, so $F_1 \sim 1$ is a good approximation. 
Fig.~\ref{boost13a} (left panel) shows $T_2\,(F_2-1) / p'_t \cos(2 \Delta \phi_r) \sim \Delta y_{t2} = 0.2$. Additional azimuth structure due to higher-order terms in the exponential is substantial at larger $y_t$.
%%%%%%%%%%
%%%%%%%%%%
The full integral over factors $F_1$ and $F_2$ is
\bea \label{fint}
 \frac{1}{2\pi} \int_{-\pi}^{\pi} \hspace{-.1in} d\phi\, F_1(\phi)\, F_2(\phi)\,\cos(2 \Delta \phi_r)  
\hspace{-.05in } &\equiv& \hspace{-.05in} \frac{\Delta y_{t2}\, p'_t\,f(y_t)}{2T_2},
\eea
defining $f(y_t;\Delta y_{t0},\Delta y_{t2})$ as an $O(1)$ correction factor plotted in Fig.~\ref{boost13a} (right panel) for a particular combination $(\Delta y_{t0},\Delta y_{t2})$. $f(y_t)$ is closer to 1 the smaller is $\Delta y_{t2}/\Delta y_{t0}$.

Combining factors we obtain
\bea \label{combfac}
\frac{V_2(y_t;\Delta y_{t0},\Delta y_{t2})}{2\pi} \hspace{-.04in} &=&  \hspace{-.04in}  \frac{p'_t}{2T_2} f(y_t)\, \Delta y_{t2}\, \rho_2(y_t;\Delta y_{t0}).
\eea
In Fig.~\ref{boost14ab} (right panels) the dashed curves through points represent $p'_t/2\cdot f(y_t)\, \rho_2(y_t;\Delta y_{t0})$, which agree well with the direct integrals (points) except in the region of negative values near $\Delta y_{t0}$ where accurate comparisons with data are not possible. The dash-dot curves represent $p'_t/2\cdot \rho_2(y_t;\Delta y_{t0})$ which does not include factor $f(y_t;\Delta y_{t2},\Delta y_{t0})$. Small deviations from the exact integral (dashed curves and points) due to omission of factor $f(y_t)$ depend on ratio $\Delta y_{t2} / \Delta y_{t0}$ as noted.

Because $V_2/2\pi = \rho_0\, v_2$, $v_2(p_t)$ can be expressed as
\bea \label{v2def}
v_2(p_t) &=&  \frac{p'_t }{2T_2} \frac{ f(y_t)\,\Delta y_{t2} \,\rho_2(y_t; \Delta y_{t0},T_2,n_2)}{\rho_0(y_t;T_0,n_0)} 
\eea 
Given Eq.~(\ref{v2def}) we can reconstruct at least the shape of quadrupole spectrum $\rho_2(y_t;\Delta y_{t0})$ from measured $v_2(p_t)$ data.  Although the derivation is based on an exponential form for $\rho_2$, the procedure can be applied to the more general form of a L\'evy distribution within the limited $p_t$ range relevant to quadrupole and soft components ($\leq 2$ GeV/c in the boosted frame).

\subsection{Obtaining $\rho_2(y_t;\Delta y_{t0})$ from measured $v_2$ data}

The quadrupole spectrum is best related to measured quantities with the equation
\bea \label{stuff}
\rho_0(y_t)\, \frac{v_2(y_t)}{p_t} &=& \left\{\frac{p'_t}{p_t\, \gamma_t(1 - \beta_t)}\right\}\,  \left\{\frac{\gamma_t(1 - \beta_t)}{2T_2}\right\} \times \\ \nonumber
&&f(y_t;\Delta y_{t0},\Delta y_{t2})\,\Delta y_{t2}\,  \rho_2(y_t;\Delta y_{t0}).
\eea
Quantities on the LHS are measured experimentally. $\rho_2(y_t;\Delta y_{t0})$ on the RHS is the sought-after quadrupole spectrum. The common monopole boost $\Delta y_{t0}$ and $T_2$ for each hadron species can be estimated accurately from the $\rho_2(y_t;\Delta y_{t0})$ spectrum common left edge and shape. As shown in Fig.~\ref{boost3} (right panel) $p'_t / \{p_t \gamma_{t}(1 - \beta_{t})\}$ is determined only by $\Delta y_{t0}$ and deviates from unity only near that point. The numerator of the second factor is also determined by $\Delta y_{t0}$. Thus, all factors on the {\em first line} of the RHS and the shape of $\rho_2(y_t;\Delta y_{t0})$ are determined by data on the LHS.

In the second line of Eq.~(\ref{stuff}) RHS there is an ambiguity in the product of $\Delta y_{t2}$ and the $\rho_2(y_t;\Delta y_{t0})$ amplitude. Comparison of the inferred quadrupole $\rho_2(y_t;\Delta y_{t0})$ spectrum shape, especially the leading edge of the spectrum, with measured azimuth-averaged spectrum $\rho_0$ for each hadron species may place a lower limit on $\Delta y_{t2}$. The upper limit $\Delta y_{t2} \leq \Delta y_{t0}$ assumes positive-definite transverse boosts. The two limits establish an allowed range for quadrupole spectrum integral $n_{ch2}$. $\Delta y_{t2}$ should be common to all  hadron species emitted from a boosted hadron source, possibly reducing systematic uncertainty. 

%The estimate of $f(y_t;\Delta y_{t0},\Delta y_{t2}) \sim 1$ can be refined iteratively. 

\subsection{Analysis summary}

To summarize, given measurements of $v_2(p_t)$ the quadrupole spectrum is determined with minimal systematic uncertainty by the following steps for each hadron species
\begin{enumerate}
\item Parameterize single-particle spectrum $\rho_0(y_{t\pi})$; obtain the value of $\rho_0(y_{t\pi})$ for each $v_2$ datum
\item Calculate and plot $\rho_0(y_{t\pi})\, v_2(p_t) / p_t(\text{lab frame})$
\item Model $ \rho_2(y_t,\Delta y_{t0})$ by a boosted L\'evy distribution
\item Use the model to plot the product of the first and last factors of Eq.~(\ref{stuff}) RHS on $y_{t\pi}$
\item Compare 4) with 2) to determine monopole boost $\Delta y_{t0}$ and temperature $T_2$ plus L\'evy $n_2$
\item Obtain product $f(y_t)\, \Delta y_{t2}\, \rho_2(y_t,\Delta y_{t0})$ from Eq.~(\ref{stuff})
\item Compare the inferred $\rho_2(y_t,\Delta y_{t0})$ shape with single-particle spectra to obtain an upper bound on the $\rho_2(y_t,\Delta y_{t0})$ amplitude $\Leftrightarrow$ lower bound on $\Delta y_{t2}$
\item Iterate $\Delta y_{t2}$ to optimize $f(y_t;\Delta y_{t0},\Delta y_{t2})$
\item Obtain corrected $\rho_2(y_t,\Delta y_{t0})$ 
\end{enumerate}
Step 7) should include comparison of the  approximate centrality variation of $\rho_2(y_t,\Delta y_{t0})$ with the measured centrality variations of the single-particle spectrum components on $y_{t\pi}$~\cite{2comp} to tighten constraints.

%%%%%%%%%%
\section{Quadrupole obtained from data} \label{quad-dat}

The analysis procedure can be illustrated with the data shown in Fig.~\ref{boost1a} following the steps described in the previous section. The procedure is intended to minimize model assumptions and systematic errors. Step 1, defining single-particle spectrum parameterizations, was described in Sec.~\ref{singlespec}.

\subsection{Forming the LHS of Eq.~(\ref{stuff})}

Steps 1 and 2 of the analysis produce the LHS of Eq.~(\ref{stuff}) from $v_2(p_t)$ data. Fig.~\ref{boost1a} (left panel) shows the original $v_2$ data in a form which provides little direct indication of the underlying physics. The left-most measured Lambda point (not plotted) is negative, the reason now apparent from the discussion in Sec.~\ref{negproton}.  Fig.~\ref{boost1a} (right panel) shows $v_2(p_t) / p_t$ which hints at a simple boost phenomenon. The common left edge provides an initial estimate of $\Delta y_{t0}$.

%%%%%%%%%%
 \begin{figure}[h]
  \includegraphics[width=3.3in,height=3.3in]{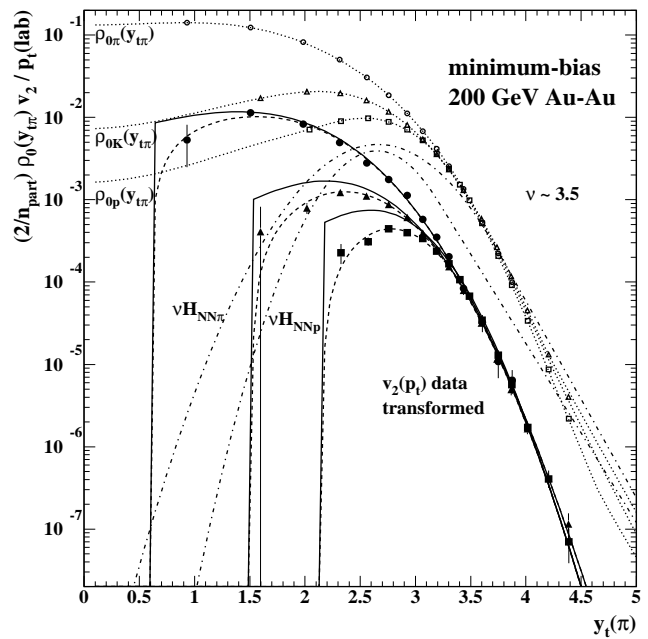}
\caption{\label{boost1bb}
Formation of quadrupole spectra from $v_2(p_t)$ data and single-particle spectra -- measured quantities combined with full-spectrum two-component parameterizations. The open symbols are the values of $\rho_0(y_t)$ used for the conversion. The solid symbols are the resulting approximations to quadrupole spectra. The dashed curves are from the present analysis. The solid curves result from removing a $p'_t / p_t$ factor. The dash-dotted curves are hard-component models from~\cite{2comp}.
} 
%boost1b
 \end{figure}
%%%%%%%%%%

Fig.~\ref{boost1bb} illustrates how to match $\rho_0(y_{t\pi})$ parameterizations to $v_2(p_t) / p_t$ data (step 1). Single-particle spectra in the form $2/n_{part} \cdot \rho_0(y_{t\pi})$ for three hadron species (protons and pions from~\cite{2comp}, kaons interpolated) are given by the dotted curves [defined in Eq.~(\ref{summ}) and discussed in Sec.~\ref{specmodels}]. The open symbols show the specific values of $\rho_0$ for each $v_2$ datum and hadron species. The solid symbols show the corresponding values of $2/n_{part} \cdot \rho_0(y_{t\pi})\, v_2(p_t) / p_t$ (step 2). 

The dashed curves show the result of steps 3 and 4---modeling the data with boosted soft component $S'_{NN}$ (L\'evy distribution). The solid curves show the RHS of Eq.~(\ref{stuff}) with the first bracket replaced by $2/n_{part}$  to form $\gamma_t(1 - \beta_t)/2T_2 \cdot \, f(y_t)\, 2/n_{part}\, \Delta y_{t2}\, \rho_2(y_t)$. The minimum-bias data used in this analysis correspond to mean participant pathlength $\nu \sim 3.5$ as noted in the figure. In the product $\rho_{0X}(y_t)\cdot v_{2Y}(y_t)$ of Fig.~\ref{boost1bb} the correspondence proton $\approx$ Lambda is made for $v_2(p_t)$ data to estimate proton quadrupole spectra.

Subsequent spectrum interpretation invokes the relation of the three quadrupole components to corresponding hard components on $y_{t\pi}$ (dash-dotted curves $H_{NNX}$ in Fig.~\ref{boost1bb}). The hard components for p and K (all hard-component modes are at $y_t(\pi) \sim 2.7$) strongly overlap the corresponding quadrupole components, but that for pions does not. Such structural details may explain the variation of $v_2(p_t)$ distributions with mass in relation to so-called constituent-quark scaling.

\subsection{Quadrupole spectrum and soft component} \label{trial}

\begin{figure}[h]
  \includegraphics[width=3.3in,height=3.3in]{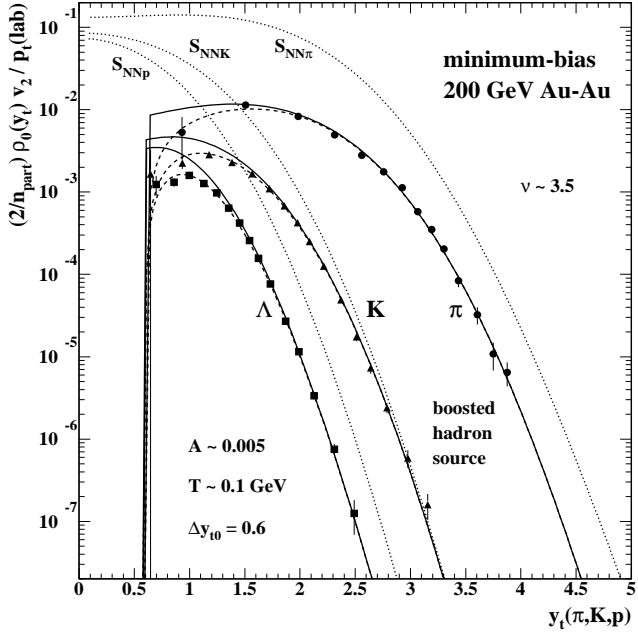}
\caption{\label{boost1ee}
Spectra from Fig.~\ref{boost1bb} transformed to proper $y_t$  for each hadron species. The dotted curves are soft components from respective single-particle spectra for comparison. The prominent feature is the common edge at $y_t \sim 0.6$, implying that the three hadron species originate from a common boosted source. The hadron abundances and spectrum shapes are the same as the single-particle spectrum soft components.
%Fit to data is sofpyxs; soft pion component is sfpiyax. 
 } 
%boost1e
 \end{figure}

Fig.~\ref{boost1ee} shows data (solid points) from Fig.~\ref{boost1bb} transformed to $y_t(\pi,K,p)$ (proper $y_t$ for each hadron species) with the appropriate Jacobians.  The common left edge reveals monopole boost $\Delta y_{t0} \simeq 0.6$. From Eq.~(\ref{stuff}) the {\em form} of the data is $\propto p'_t / p_t\, \cdot f(y_t) \cdot \rho_2(y_t;\Delta y_{t0})$, the last factor being the quadrupole spectrum.

%%%%%%%%%%
%%%%%%%%%%

The quadrupole spectrum for each hadron species can be modeled with the same form of L\'evy distribution used for the soft component of the single-particle spectrum. Also plotted in Fig.~\ref{boost1ee} are soft components $S_{NNX}(y_t)$ from the single-particle spectra for three hadron species (dotted curves). The dashed curves through data points are $A/T_2\cdot  p'_t / p_t\, \gamma_t(1 - \beta_t)\cdot S'_{NN}(y_t - \Delta y_{t0};T_2,n_2)$, with factor $A$ and monopole boost $\Delta y_{t0}$ common to the three species. L\'evy $T_2$ and $n_2$ parameters have been optimized for each quadrupole spectrum. The factors are $A / T_2 \sim 0.005 / (0.1~\text{GeV})  \sim 1/20~\text{GeV}^{-1}$. The description of data is good. The solid curves are the same but with factor $p'_t / p_t\gamma_t(1-\beta_t)$ (Fig.~\ref{boost13a}, right panel) removed, revealing the undistorted shapes of $\rho_2(y_t,\Delta y_{t0})$. Comparison with the single-particle spectra (dotted curves) reveals the similarities of the single-particle soft and quadrupole hadron sources.

%%%%%%%%%%%%%%%%
\section{The structure of ${\bf v_2}$} \label{structure}

In Sec.~\ref{quadcomp} $v_2$ was factored, and in Sec.~\ref{quad-dat} $v_2$ was represented by the combination of a boosted soft component $S'_{NN}$ and two-component single-particle spectrum ${2}/{n_{part}}\cdot  \rho_0$. The full expression with $V_2 / 2\pi = \rho_0\, v_2$ is 
\bea \label{v2struct}
\frac{2}{n_{part}} \frac{V_2(y_t)}{2\pi} \hspace{-.0in}&=& \hspace{-.0in}  \frac{p'_t }{\gamma_t(1 - \beta_t)} \frac{ A }{T_2}S'_{NN}(y_t - \Delta y_{t0};T_2) \\ \nonumber
 &=&\left\{\frac{p'_t}{ \gamma_t(1 - \beta_t)}\right\}\,  \left\{\frac{\gamma_t(1 - \beta_t)}{2T_2}\right\} \times \\ \nonumber
&&  \hspace{-.2in} f(y_t;\Delta y_{t0},\Delta y_{t2})\,\frac{2}{n_{part}}\, \Delta y_{t2}\, \rho_2(y_t;\Delta y_{t0}),
\eea
where $2/n_{part} \cdot \rho_2$ has been modeled by $A\, S'_{NN}(y_t - \Delta y_{t0};T_2,n_2)$, a boosted soft component (L\'evy distribution) with reduced amplitude and modified shape parameters ($T,n$). 

The structure of $v_2(p_t)$ can be understood entirely in terms of two $y_t$-dependent factors---$p'_t$ and the spectrum ratio.  The two-component spectrum model is ${2}/{n_{part}}\cdot  \rho_0 = S_{NN}(y_t;T_0) + \nu\,  H_{AA}(y_t,\nu)$. The shape of the single-particle spectrum in the $v_2$ denominator varies strongly with centrality ($\nu$) due to evolution of the hard component~\cite{2comp}.

\subsection{${\bf p'_t =  p_t}$ in the boost frame}

\begin{figure}[h]
   \includegraphics[width=1.65in,height=1.65in]{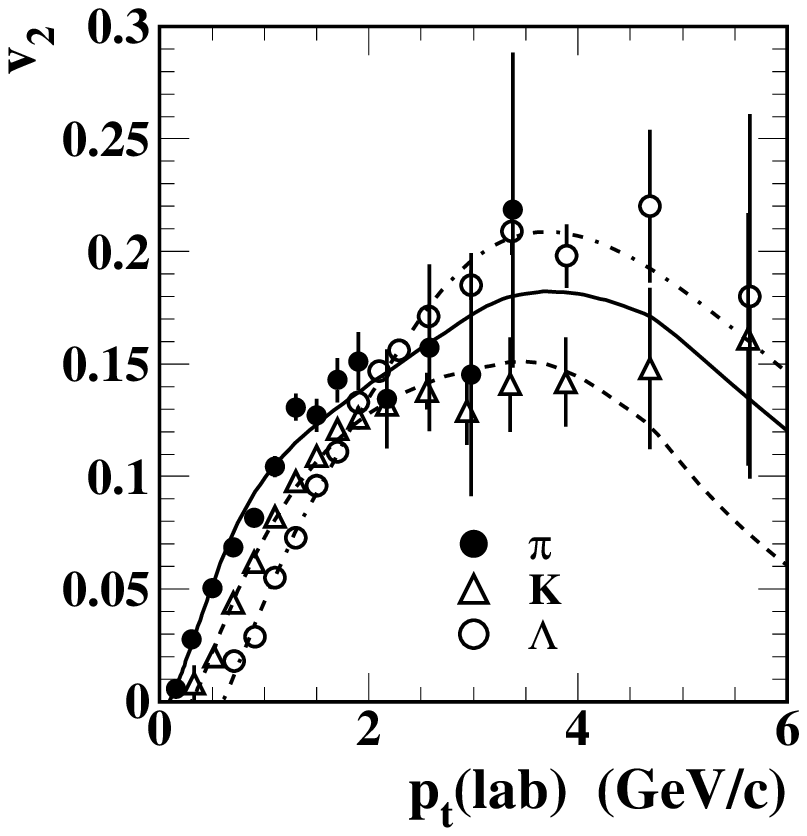}
\  \includegraphics[width=1.65in,height=1.65in]{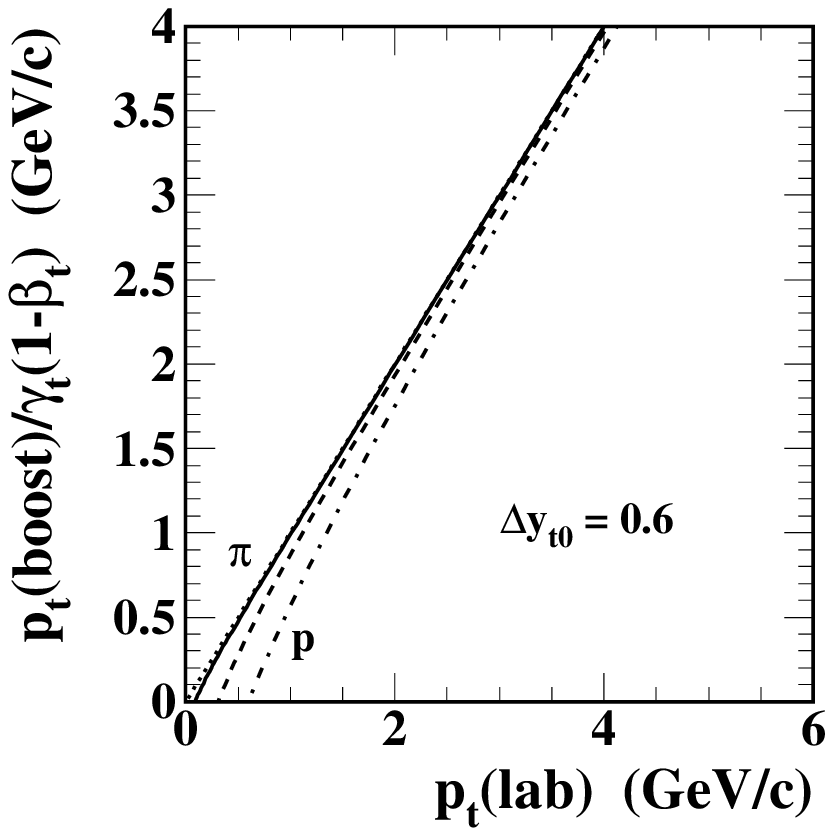}
 \caption{\label{boost3cbx}
Left panel:   $v_2(p_t)$ data from Fig.~\ref{boost1a} (left panel) are repeated for comparison. The curves through data are from the present analysis.
Right panel: Relation between $p_t$ in the boost frame and lab frame for three hadron masses. Intercepts on the abscissa are at $p_{t0}$ values defined in the text. 
 } 
%boost1a, boost3c
 \end{figure}

Fig.~\ref{boost3cbx} (left panel) shows $v_2(p_t)$ data with mass ordering at small $p_t$ attributed  to hydrodynamic flow~\cite{keystone,kolb1,kolb2}. The mass dependence of $v_2(p_t)$ at small $p_t$ is determined entirely by $p_t$ in a frame boosted on $y_t$ by $\Delta y_t$. $p_t$ in the boost frame is defined in the lab frame by
\bea
p'_t &\equiv&   m_0\, \sinh(y_t - \Delta y_{t}) =\gamma_{t}\, (p_t - \beta_{t}\, m_t) \\ \nonumber
&=& m_t\, \gamma_{t} \{  \tanh(y_t) - \tanh(\Delta y_{t})  \} .
\eea
$p'_t$ {\em vs} $p_t$ in the lab frame is shown in Fig.~\ref{boost3cbx} (right panel). As noted, the curve intercepts are located at $p_{t0} = m_0\sinh(\Delta y_{t})$. Comparing the two panels it is apparent that the ``mass ordering'' of $v_2$ attributed to hydrodynamics is produced entirely by the kinematic relation between $p'_t$ and $p_t$ determined by monopole boost $\Delta y_{t0}$. The mass ordering alone does not determine what physical mechanism caused the boost.

%%%%%%%%%%
%%%%%%%%%%

\subsection{Spectrum ratios} \label{specrat}

\begin{figure}[h]
  \includegraphics[width=1.65in,height=1.65in]{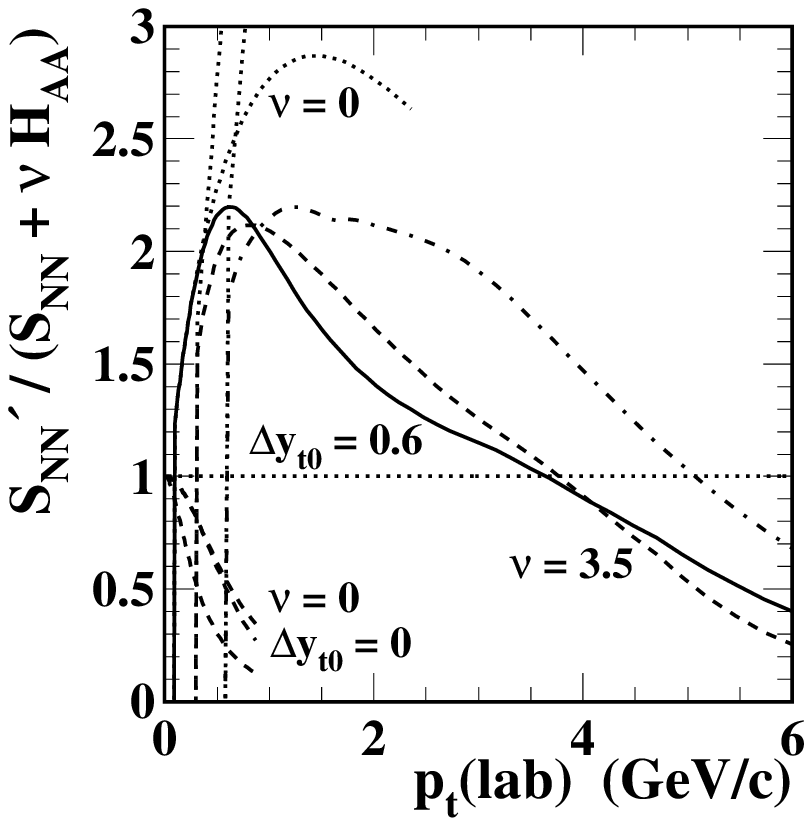}
  \includegraphics[width=1.65in,height=1.65in]{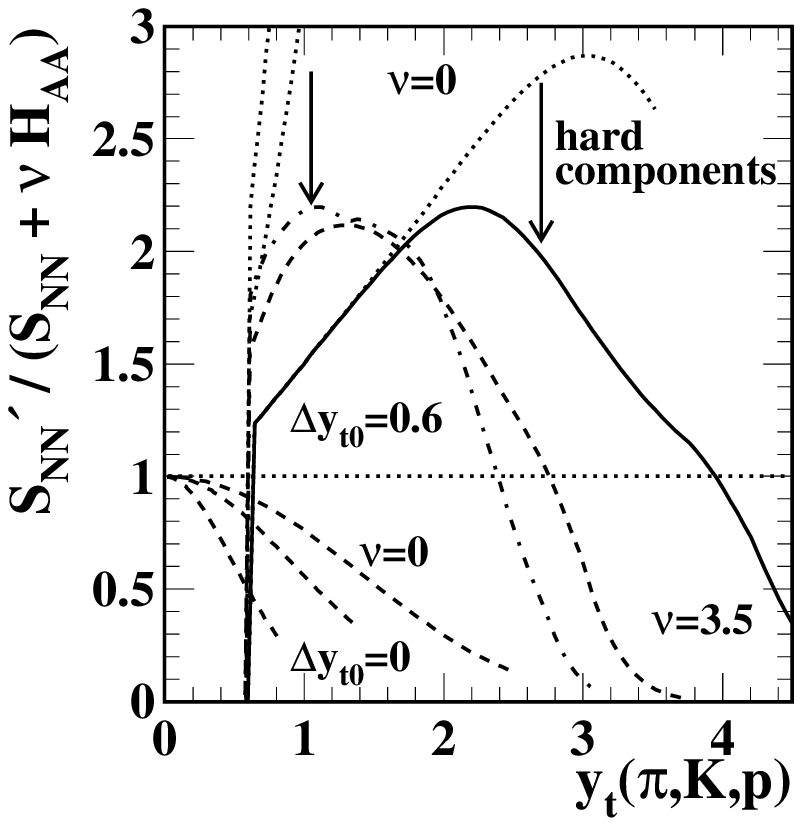}
\caption{\label{boost18ab}
Ratio of quadrupole (boosted soft component) spectra to single-particle spectra plotted on $p_t$ (left panel) and proper $y_t$ for each hadron species (right panel; solid, dashed, dash-dot) for minimum-bias Au-Au collisions ($\nu \sim 3.5$). Spectrum ratios with no hard component (dotted, $\nu = 0$) are relevant to a hydro description. Ratios for no monopole boost (dashed, $\Delta y_{t0} = 0$) are also plotted. The dominant role of the hard component in $v_2(p_t)$ is most evident in the right panel.
%Fit to data is sofpyxs; soft pion component is sfpiyax. 
 } 
%boost18b, boost18a 
 \end{figure}

Variation of $v_2(p_t)$ {\em relative to} the $p'_t$ trend (especially above 0.5 GeV/c) is determined by ratios of quadrupole to single-particle spectra. Fig.~\ref{boost18ab} shows spectrum ratios
\bea \label{ratio}
\frac{S'_{NN}(y_t - \Delta y_{t0};T_2)}{S_{NN}(y_t;T_0) + \nu\,  H_{AA}(y_t,\nu)} \propto \frac{\rho_2}{\rho_0}
\eea
in two plotting formats for three hadron species (pions -- solid, kaons -- dashed, Lambdas -- dash-dot). The numerator of Eq.~(\ref{ratio}) appears in Fig.~\ref{boost1bb} (plotted on pion $y_t$) as solid curves in the form  $A / T_2\cdot S'_{NN}(y_t - \Delta y_{t0};T_2)$ [$\propto$ quadrupole spectra $\rho_2(y_t)$]. The denominator appears in that figure as dotted curves [$\propto$ two-component single-particle spectra $\rho_0(y_t)$]. Those spectrum models describe $v_2(p_t)$ and spectrum data within published errors. 

%%%%%%%%%%
%%%%%%%%%%

In Fig.~\ref{boost18ab} the dashed curves at lower left represent spectrum ratios for no monopole boost ($\Delta y_{t0} = 0$) and no hard component ($\nu = 0$), which then directly relate the quadrupole boosted $S'_{NN}$ {shape} to the soft-component shape of the single-particle spectrum. The ratios are defined to obtain unit magnitude at $y_t = p_t = 0$ for those conditions.  Quadrupole spectra inferred from $v_2(p_t)$ data are significantly narrower for each hadron species, i.e. substantially ``cooler'' than the single-particle soft components ($T_2 \sim 0.7\, T_0$). The dotted curves at upper left in each panel result from ``turning on'' the quadrupole boost $\Delta y_{t2} = 0.6$ but not the hard component (minijets) in the denominator of Eq.~(\ref{ratio}) ($\nu \rightarrow 0$) to form {\em soft-reference} spectrum ratios. 

The $\nu = 3.5$ spectrum ratios (solid, dashed, dash-dot curves) include the hard component in the denominator. Deviations from the $\nu = 0$ (dotted) soft-ratio reference curves (arrows) demonstrate the {\em dominant role of the hard component} in distorting spectrum ratios, and therefore $v_2(p_t)$, over most of the $p_t/y_t$ range. The distortion is not a consequence of ``nonflow'' as defined in conventional flow analysis (discussed in Sec.~\ref{nonflow}). It is inherent in the $v_2(p_t)$ definition as a ratio and results from the minijet contribution to its denominator.

The mass systematics for the quadrupole and soft components are simplest to describe in Fig.~\ref{boost18ab} (right panel). From Eq.~(\ref{v2struct}) spectrum ratios are  $\propto v_2(p_t) / p'_t$. Fig.~\ref{boost18ab} (right panel) is therefore comparable to  $v_2 / p_t$ data in Fig.~\ref{boost1a} (right panel). Soft-component or ``thermal'' spectra ($\nu = 0$) on specific hadron $y_t$ vary approximately as $\exp\{-m_0[\cosh(y_t) - 1]/T\}$. The spectrum widths then vary as $1/m_0$, as illustrated by the dashed curves at lower left for $\Delta y_{t0} = 0$. For nonzero $\Delta y_{t0}$ ($\sim 0.6$) the same  trend holds for increases above unity of the soft-component ratios ($\nu = 0$), as illustrated by the dotted curves. Deviations of data ratios from the dotted curves (arrows) are determined by the relation between soft and hard spectrum components for different hadron species.

\subsection{Role of the hard component in ${\bf v_2(p_t)}$}

Because the modes of the hard components for three hadron species are located near $1$ GeV/c (reflecting a common underlying parton spectrum) spectrum ratios for more-massive hadrons are more affected at smaller $y_t$. The quadrupole and minijet peaks for kaons and Lambdas/protons occur at nearly the same position on $p_t$, whereas the peaks for pions are significantly separated, as in Fig.~\ref{boost1bb}.  Thus, the pion ratio is affected only above about 0.5 GeV/c, but the kaon and Lambda (proton) spectrum ratios are dominated by the hard component over all $p_t/y_t$, explaining the mass dependence of the sequence of downturns (arrows) from the dotted curves in Fig.~\ref{boost18ab} (right panel).

In effect, the hard component interacts through ratio $v_2$ with the quadrupole component and soft spectrum component. There is a complex numerical interplay between hadrons from a boosted source, longitudinal participant-nucleon fragmentation and transverse parton fragmentation. The correspondence below $y_t \sim 1.5$ arises because the boosted hadron source apparently produces hadron species in the same abundance ratios as the N-N soft component (cf. statistical model~\cite{stat1,stat2}). The main difference is a smaller slope parameter $T_2 \sim 0.7\, T_0$. This exercise demonstrates the importance of {\em interpretable} correlation measures and plotting formats. Quadrupole and minijet systematics should be studied within independent analysis contexts. In $v_2(p_t)$ they are {maximally confused}.

%%%%%%%%%%%%%%%%%%%%
\section{Quantitative comparisons}

The description of quadrupole spectra in Sec.~\ref{structure} can be related quantitatively to other spectrum features and trends. The goal should be the simplest and most comprehensive description of all single-particle and correlation structures.

\subsection{Relation to the soft-component spectrum}

From Fig.~\ref{boost1bb} we can infer that
\bea
\hspace{-.05in}\frac{2}{n_{part}} \rho_{0X} \frac{v_{2X}(p_t)}{p_t} \hspace{-.05in} &\approx&\hspace{-.05in} \frac{0.005}{T_2} S'_{NNX}(y_t \hspace{-.02in}-\hspace{-.02in} \Delta y_{t0};T_2,n_2)
\eea
for each hadron species X. From Eq.~(\ref{stuff}) we then have 
\bea \label{quadspecx}
\frac{2}{n_{part}}  \Delta y_{t2}\,\rho_{2X} &\simeq& \frac{2\cdot 0.005}{\gamma_t(1-\beta_t)}\, S'_{NNX}(y_t - \Delta y_{t0}) \\ \nonumber
 &\simeq& 0.016\, S'_{NNX}(y_t - \Delta y_{t0}),
\eea
since $\gamma_t(1-\beta_t) \sim 0.6$ and $T_2 \sim 0.1$ GeV for minimum-bias data.  Thus, N-N soft particle production from a common boosted source describes the quadrupole in A-A minimum-bias collisions.
 
\subsection{Relation to soft and hard hadron yields}

Minimum-bias spectrum data do not provide information about centrality dependence. In~\cite{gluequad} the centrality trend for $p_t$-integrated $v_2$ was inferred from $v_2\{4\}$ data
\bea
\frac{1}{\rho_{0}}\,\frac{V_2^2}{(2\pi)^2} &=& 0.0045\, \epsilon_\text{optical}^2 \, n_{bin} \\ \nonumber
\frac{2}{n_{part}}\frac{V_2}{2\pi\, \epsilon_\text{optical}} &=&\sqrt{0.0045\,\nu \frac{2}{n_{part}}\frac{n_{ch}}{2\pi} }.
%\\ \nonumber
%\left\{\frac{2}{n_{part}}\frac{V_2}{2\pi\,\epsilon} \right\}^2&\simeq& 0.0045 \, \nu \, n_{NN} / 2\pi
\eea
The per-participant hadron {\em quadrupole density squared} $\propto \nu$, whereas for minijets the per-participant hadron {\em fragment density} $\propto \nu$, possibly revealing the difference between quadrupole radiation and parton scattering~\cite{gluequad}. 

Integrating Eq.~(\ref{combfac}),  with $n_{ch2}$ the integral of quadrupole spectrum $\rho_2$ over $p_t$ and {\em one unit of rapidity}, gives
\bea
\frac{2}{n_{part}}\frac{V_2}{2\pi\, \epsilon} &\simeq& \frac{\bar p'_t}{2T_2} \frac{1}{\epsilon}\, \frac{2}{n_{part}}\frac{\Delta y_{t2}\,n_{ch2}}{2\pi},
\eea
where $\Delta y_{t2}\, n_{ch2}$ is the effective number of quadrupole hadrons in one unit of rapidity. Given the $v_2(p_t)$ data $n_{ch2}$ could be a large number with weak boost modulation $\Delta y_{t2}$ or a small number with strong modulation. Invoking $2/n_{part} \cdot n_{ch} \simeq n_{NN}(1 + 0.1[\nu - 1]) \sim n_{NN} = 2.5$ at 200 GeV we have
\bea \label{quadcent}
\frac{2}{n_{part}}\frac{\Delta y_{t2}\,n_{ch2}}{2\pi} &\approx& {\epsilon} \frac{2T_2}{\bar p'_t} \, \sqrt{0.0045\, \nu\, n_{NN}/2\pi} \\ \nonumber
 \frac{2}{n_{part}} \frac{\Delta y_{t2}\,n_{ch2}}{n_{NN}} &\sim& 0.028\, \epsilon\, \sqrt{\nu}  \\ \nonumber
 &\sim& 0.016,
\eea
where the last line applies to the minimum-bias case, with $\nu \sim 3.5$ and $\epsilon \sim 0.3$.
%\bea
%\frac{2}{n_{part}} \frac{\hat n_2}{n_{NN}} &\sim& 0.03,  
%\eea
From Eq.~(\ref{quadspecx}) with $n_\text{soft} \sim n_{NN} $
\bea
\frac{2}{n_{part}} \frac{\Delta y_{t2}\,n_{ch2}}{n_{NN}} &\sim& 0.016
\eea
for minimum-bias collisions, which is consistent. An independent method is required to place limits on the absolute multiplicity $n_{ch2}$ of the quadrupole (cf. Sec.~\ref{quadspec}). For comparison, the full-spectrum and hard-component integrals relative to the N-N multiplicity are
\bea
\frac{2}{n_{part}}\frac{n_{ch}}{n_{NN}} &=& 1 + 0.1(\nu - 1) \\ \nonumber
\frac{2}{n_{part}}\frac{n_{hard}}{n_{NN}} &=& 0.1\nu.
\eea
Roughly, $2/n_{part}\, n_\text{hard} \sim 0.25 \nu$.

The factors in the second line of Eq.~(\ref{quadcent}) can be interpreted in terms of quadrupole emission as follows: The final-state quadrupole moment (hadron pair yield) $\propto(\Delta y_{t2}\,n_{ch2})^2$ goes as interaction length ${\nu}$ times the initial-state quadrupole moment $\propto (n_{part} / 2\cdot \epsilon_\text{opt})^2$.

\subsection{Relation to minijet pair correlations}

The hard-component spectrum yield should relate to observed changes in pair correlations associated with minijets. The factor 
\bea
\frac{2}{n_{part}}\, \sqrt{\rho_\text{ref}} &=& \frac{1}{2\pi} \frac{dn}{d\eta} = \rho_0 \sim 0.4(1+0.1[\nu-1]) 
\eea
times $\Delta \rho / \sqrt{\rho_\text{ref}}$ (per particle) gives the number of correlated hadron  pairs per {\em participant nucleon pair}.
The density of correlated fragment pairs per participant pair is therefore
\bea
\frac{2}{n_{part}} \Delta \rho(\eta_\Delta,\phi_\Delta) &=&  0.4(1+0.1[\nu-1]) \frac{ \Delta \rho}{\sqrt{\rho_\text{ref}}}.
\eea
The integral of ${ \Delta \rho}{\sqrt{\rho_\text{ref}}}$ over the same-side minijet peak increases by $6\times$ over binary-collision scaling. The increase in number of correlated pairs per participant over binary collision scaling for central collisions is therefore $1.5\cdot 6 = 9\times$. That means the number of hadron fragments per parton pair increases by $3\times$ from $\sim 2.5$ to $\sim 7.5$.

For binary collision scaling of N-N collisions the fractional increase of hadrons which are correlated minijet fragments from peripheral to central Au-Au should be 8\%. It is observed to be 25\% due to the factor $3\times$ derived above. The fraction of total hadrons which is correlated fragments is thus 25/125 = 20\%. 

The hard-component fraction of single-particle spectra is $\sim 30$\% for central Au-Au collisions. The hard component of spectra appears entirely in narrow structures for pions and protons corresponding to the same boost~\cite{2comp}. The 13\% difference is therefore intimately connected with the anomalous boost component, which is in turn connected with minijet broadening. %The relation to correlations is not yet clear.

%%%%%%%%%%
\section{Quadrupole {\em vs} Hydro Theory} \label{theorycomp}

Using this detailed description of $v_2(p_t)$ structure we can explore the relationship of the quadrupole component to  theory predictions. Hydrodynamics is tested by the quadrupole component in three ways: the source boost distribution, the apparent ``temperature'' $T_2$ of the quadrupole source and the abundances of the several hadron species produced by the source. Parameters of the spectrum soft component, such as temperature $T_0$, are obtained from measured single-particle spectra.

In Fig.~\ref{boost1a} dotted curves in the two panels represent viscous hydro calculations for pions~\cite{teaney,rom} which have been interpreted to support claims for formation of a low-viscosity medium, possibly a ``perfect liquid.'' The theory curves represent the limiting case of zero viscosity. Although the curves in the left panel {\em appear} to describe the pion data, transformation to the right panel reveals that theory and data are actually substantially different.

\subsection{Spectrum ratios}

Typical hydro calculations do not include a hard component in the denominator of $v_2(p_t)$. Thus, the spectrum ratio in Eq.~(\ref{ratio}) becomes
\bea \label{hydroeq}
\frac{S'_{NN}(y_t - \Delta y_{t0})}{S_{NN}(y_t)} &\sim& \frac{\exp\{- \gamma_t(m_t - \beta_t\, p_t)/T_2\}}{\exp\{- m_t/T_0\}} \\ \nonumber
&\sim & \exp\{ [1/T_0 - \gamma_t(1-\beta_t)/T_2]\, m_t\},
%\\ \nonumber
%&\sim&\exp\{ \sim(0 \rightarrow 0.4)\, m_t/\bar T\}
\eea
where the boosted and single-particle soft components are approximated by exponentials with possibly different temperatures, and $\gamma_t,\, \beta_t$ are determined by monopole boost $\Delta y_{t0}$. In a quantitative analysis the soft components are modeled by L\'evy distributions with possibly different indices $n$. The ratio is $\sim$ an exponential with positive constant determined by the temperature difference and the monopole boost, as in the second line.

%%%%%%%%%%
 \begin{figure}[h]
   \includegraphics[width=1.65in,height=1.65in]{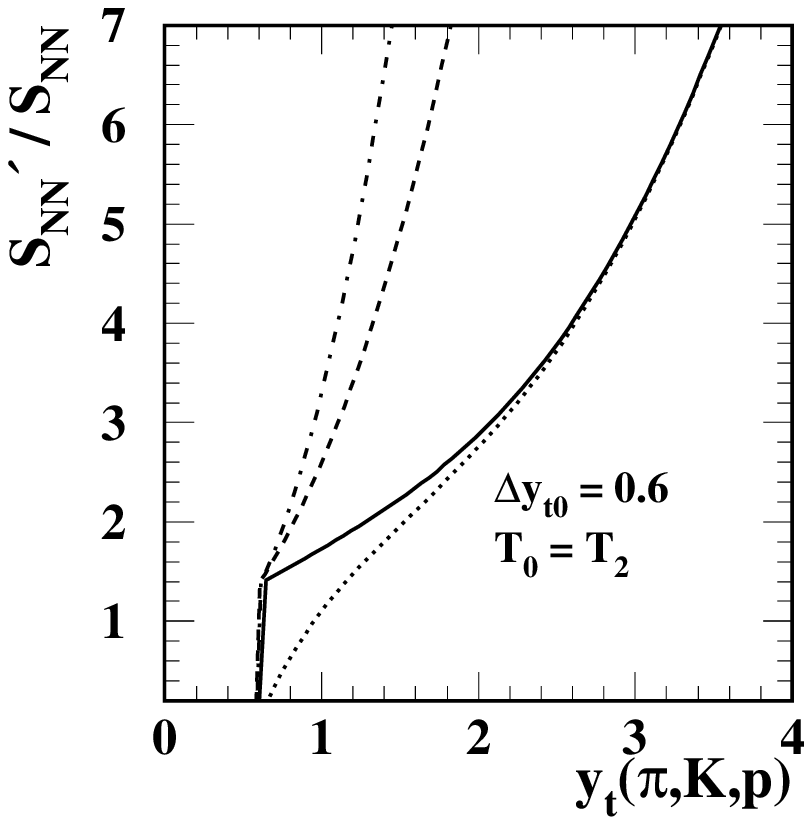}
 \includegraphics[width=1.65in,height=1.65in]{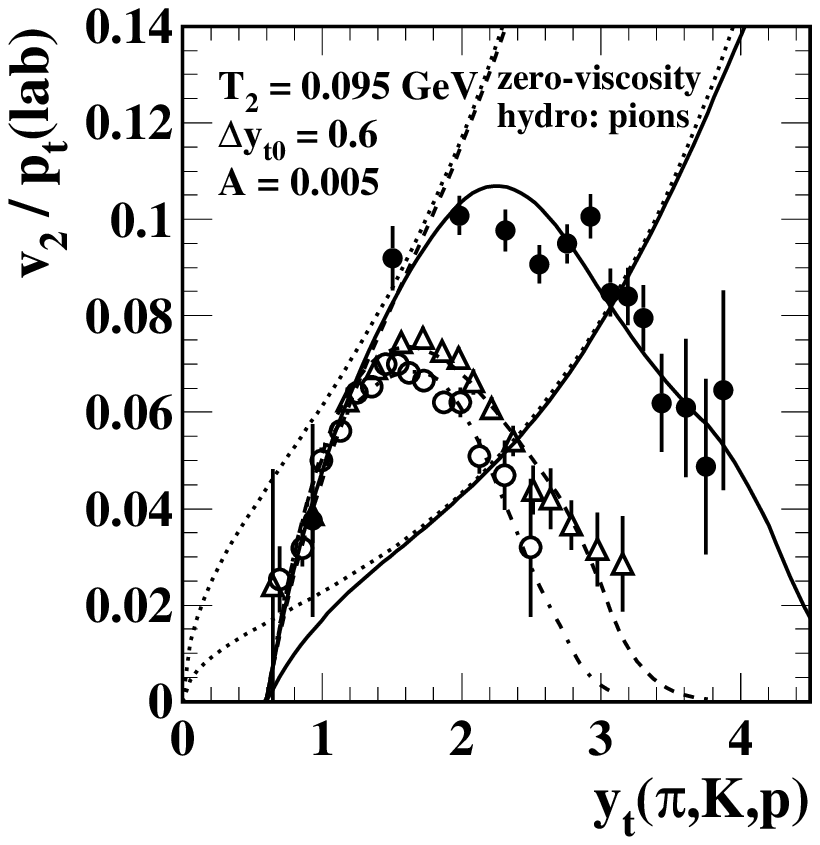}
\caption{\label{boost18cd}
Left panel: Ratio of quadrupole component (boosted soft component) to spectrum soft component (both L\'evy distributions) for three hadron species illustrating typical hydro mass dependence of the ratio. The dotted curve includes factor $p'_t/p_t \gamma_t(1-\beta_t)$ for comparison with the right panel. 
Right panel: $v_2 / p_t$ data compared to hydro theory curves as described in the text. Different boost distributions are apparent at lower left.
%Fit to data is sofpyxs; soft pion component is sfpiyax. 
 } 
%boost18c, boost19a
 \end{figure}
%%%%%%%%%%

In Fig.~\ref{boost18cd} (left panel) the solid, dashed and dash-dot curves show spectrum ratios as in Eq.~(\ref{hydroeq}) for three hadron species using spectrum soft components (denominators) which properly describe single-particle spectra (i.e., L\'evy distributions) and boosted components (numerators) with the {\em same temperatures} for simplicity and boost $\Delta y_{t0} = 0.6$ as for $v_2(p_t)$ data, showing the expected mass sequence on proper $y_t$. The ratios correspond to the dotted curves in Fig.~\ref{boost18ab} (right panel), 
but the amplitudes are adjusted so all curves start at the same initial value. The dotted curve, including additional factor $p'_r / \gamma_t(1+\beta_t)p_t$, is within a constant factor the solid curve in the right panel.

\subsection{Data comparisons with hydro theory}

In Fig.~\ref{boost18cd} (right panel) the data from Fig.~\ref{boost1a} (right panel) are repeated for comparison (the three curves through the data points are from this analysis). The lower dotted curve is the zero-viscosity hydro prediction from~\cite{teaney}. The solid curve following the hydro dotted curve at larger $y_t$  is $B\, p'_t / p_t \cdot {S'_{NN}(y_t - \Delta y_{t0})}/{S_{NN}(y_t)}$ ($\propto$ the dotted curve in the left panel). The soft-component ratio ${S'_{NN}}/{S_{NN}}$ is defined in Eq.~(\ref{hydroeq}), with $T_0 = 0.14$ GeV, $T_2 = 0.095$ GeV and $\Delta y_{t0} = 0.6$. Factor $p'_t/p_t$ has been added to incorporate the form plotted in Fig.~\ref{boost13a} (right panel) appropriate for the $v_2(p_t) / p_t$ ratio. $B$ is adjusted to match the hydro (lower dotted) curve at larger $y_t$. Agreement of the shapes is good except near the origin where the boost distributions differ.

The dashed and upper dotted curves in the right panel are $2.7\times$ the solid and hydro curves. The dashed curve describes the data for three masses well in the smaller-$y_t$ region where the hard component does not dominate the variation, as expected from Fig.~\ref{boost18ab} (right panel). The zero-viscosity hydro curve for pions from~\cite{teaney} thus underpredicts the $v_2$ magnitude by $2.7\times$ in $p_t \leq 0.5$ GeV/c ($y_{t\pi} < 2$). This exercise is intended to demonstrate how hydro theory can be better tested in the plotting format of Fig.~\ref{boost18cd}, in particular the predicted boost distribution. The validity of specific hydro theories is the subject of subsequent analysis.

\subsection{Comparison of boost distributions}

In Fig.~\ref{boost18cd} (right panel) the structure in $y_t \leq 1.5$ is possibly the first {\em direct} comparison of boost distributions  from data and hydro theory.  Boost comparisons provide essential tests of hydro and the expanding bulk medium scenario for heavy ion collisions. Boost details are strongly suppressed in plots of $v_2(p_t)$ {\em vs} $p_t$. The data for kaons and Lambdas in the form $v_2(p_t) / p_t$ on proper $y_t$ clearly contradict the hydro prediction below $y_t \sim 1.5$ where the boost distribution dominates. 

The data indicate a narrow boost distribution centered at $\Delta y_{t0} = 0.6$. The hydro prediction suggests a broad distribution starting at $y_t = 0$ and roughly consistent with Hubble expansion of a bulk medium. An essential requirement for any theory of the quadrupole component is an explicit boost distribution compared with accurate data. Inferences of small (or any) viscosity from comparisons of $v_2(p_t)$ data with the lower dotted curve~\cite{teaney} are not justified.

In Fig.~\ref{boost18cd} (right panel)  there is the suggestion of real ``scaling'' derived from a common boosted hadron source, the correspondence of the three hadron species in this plotting format below $y_t \sim 1.5$.  Above that point the curves deviate from the hydro hypothesis by large factors determined by spectrum hard components (parton scattering and fragmentation). Each hadron species deviates from the universal hydro curve at a point depending on its mass, revealing interaction of soft components with the universality of the underlying parton spectrum on $p_t$.

%%%%%%%%%
\section{Quadrupole model uncertainties}

The quadrupole spectra in Figs.~\ref{boost1bb} and \ref{boost1ee}, the main results of this analysis, were obtained as simple combinations of previously measured data. As such, uncertainties indicated by error bars are propagated from the original published errors, but spectrum parameters inferred from those data possess unique uncertainties to be estimated.

The common left edges in Fig.~\ref{boost1ee} taken together determine $\Delta y_{t_0} = 0.6 \pm 0.05$. The boost distribution appears to be narrow (r.m.s $< 0.1$) even though this is a minimum-bias centrality sample, but the data are too sparse in that region to provide a better width estimate.

The spectrum shapes near the left edges determine $T_2 \simeq 0.1 \pm 0.005$ GeV ($\sim0.09$ GeV for pions and $\sim0.11$ GeV for protons). The shapes further out on the tails of the distributions determine L\'evy exponents $n_2 \sim 15$, but the shapes are also influenced by $f(y_t;\Delta y_{t0},\Delta y_{t2}) \geq 1$. Since the spectra in Fig.~\ref{boost1ee} are uncorrected for that $O(1)$ factor the $T_2$ estimates should be taken as {\em upper limits} and the L\'evy index $n_2$ estimates as {\em lower limits}.

The largest uncertainties apply to the estimates of absolute quadrupole yields. In Fig.~\ref{boost1ee} the quadrupole amplitudes at spectrum left edges determine the {\em relative} total yields, which correspond well to the single-particle spectrum soft-component relative yields (dotted curves in that figure). Yield uncertainties from $f(y_t)$, which mainly affects the spectrum tails at larger $y_t$, are small. 

The quadrupole spectra contain a common factor $\Delta y_{t2}$ which is the major source of uncertainty in estimating the total spectrum yields, as discussed in Sec.~\ref{quadspec}. The uncertainty in the absolute yields is less than a factor $2\times$, sufficient to determine that the quadrupole component is at most a small fraction of the total particle yield.

Quadrupole $y_t$ spectra observed directly are simple, described by a few parameters and very similar in shape to single-particle spectrum soft components, albeit boosted on $y_t$. However, when coupled to two-component spectra {\em via} ratio $v_2(p_t)$ ``elliptic flow'' the data become arbitrarily complex and essentially uninterpretable.

%%%%%%%%%%%%%%%%%%%%%%%%%%%
\section{quadrupole absolute yields}  \label{quadspec}

The quadrupole absolute yield $n_{ch2}$ (in one unit of rapidity) can provide definitive model tests.  But from $v_2(p_t)$ data alone there remains an ambiguity in the product $\Delta y_{t2}\, n_{ch2}$. The ambiguity is reduced by the edge of the quadrupole spectrum near $y_t \sim \Delta y_{t0}$, evident in Fig.~\ref{boost14ab} (right panels), but accurate data in that region are difficult to obtain.

%, and an alternative strategy is required.

The single-particle spectrum hard component, described as minimum-bias parton fragmentation to minijets, suggests a solution. The hard component was isolated by a combination of techniques: correlation analysis of several hadron charge-sign combinations~\cite{ptscale,edep}, $n_{ch}$ dependence of p-p spectra~\cite{ppprd}, $\nu$ dependence of Au-Au spectra~\cite{2comp} and comparisons with the  systematics of fragmentation functions from $e^+$-$e^-$ collisions~\cite{lepmini}. 

In this section comparisons of spectrum shapes and centrality trends for integrated yields are combined to estimate the absolute quadrupole yield. Quadrupole spectra extracted from $v_2(p_t)$ data are compared to single-particle Au-Au spectra, and centrality trends of $p_t$-integrated quadrupole data are compared with those of integrated single-particle spectrum structures. 

\subsection{Lower limits from the soft-component model}

\begin{figure}[h]
  \includegraphics[width=3.3in,height=3.3in]{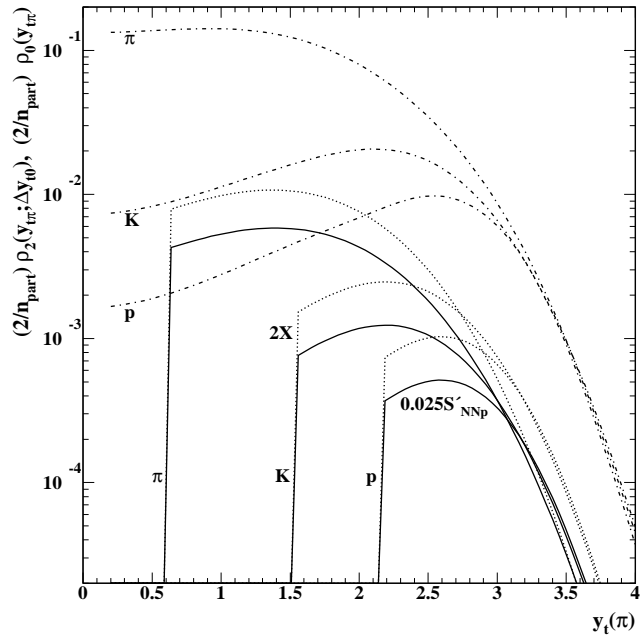}
\caption{\label{boost1h}
Lower (solid) and upper (dotted) limits on quadrupole spectra, the latter obtained by direct comparison to single-particle spectra (dash-dot curves).
 } 
%boost1h 
 \end{figure}

The boosted soft-component model of quadrupole data (dashed curves) in Fig.~\ref{boost1bb} can be used to provide a lower limit to quadrupole spectra. The dashed model curves are defined according to Eq.~(\ref{v2struct}) by
\bea \label{eqxx}
\frac{2}{n_{part}} \frac{\rho_0\, v_2}{p_t}  \hspace{-.05in} &=& \hspace{-.05in}\left\{\frac{p'_t}{p_t\, \gamma_t(1-\beta_t)} \right\}\frac{ A}{T_2} S'_{NN}(y_t - \Delta y_{t0})
\eea
with $A \sim 0.005$ and $A / T_2 \sim 0.05$. From Eq.~(\ref{v2def})
\bea
\rho_0\, v_2 = \frac{V_2(y_t)}{ 2\pi} &=& \frac{p'_t}{2T_2}\, f(y_t)\, \Delta y_{t2}\, \rho_2(y_t;\Delta y_{t0}).
\eea
A requirement of positive-definite boosts implies $\Delta y_{t2} \leq \Delta y_{t0} \approx 0.6$ and we obtain
\bea \label{eqyy}
\frac{2}{n_{part}} \rho_2 &\approx& \frac{ 2A\, S'_{NN}}{\gamma_t(1-\beta_t)\,\Delta y_{t2}} \geq 0.025 S'_{NN}.
\eea
The lower limits are represented by the solid curves in Fig.~\ref{boost1h}, which 
%The kinematic factor in curly brackets in Eq.~(\ref{eqxx}) is omitted. 
 are about $0.5\times$ the solid curves in  Fig.~\ref{boost1bb} because $0.025T_2 / A \sim 0.5$.

%%%%%%%%%%
%%%%%%%%%%

\subsection{Upper limits from single-particle spectra}

Loose upper limits on  $\rho_2(y_{t\pi};\Delta y_{t0})$ can be estimated by direct comparison with the full single-particle spectra.  Factor $f(y_t)$ is ignored, since only the most prominent spectrum aspects at smaller $y_t$ matter, {especially the edges of the boosted distributions}. In Fig.~\ref{boost1h} the rough upper limits on quadrupole spectra (dotted curves) are determined by the condition that they not exceed 10\% of the measure single-particle spectra at any point. The upper limits are then only a factor $2\times$ the lower limits. 

\subsection{Upper limits from spectrum residuals}

\begin{figure}[h]
  \includegraphics[width=1.65in,height=1.65in]{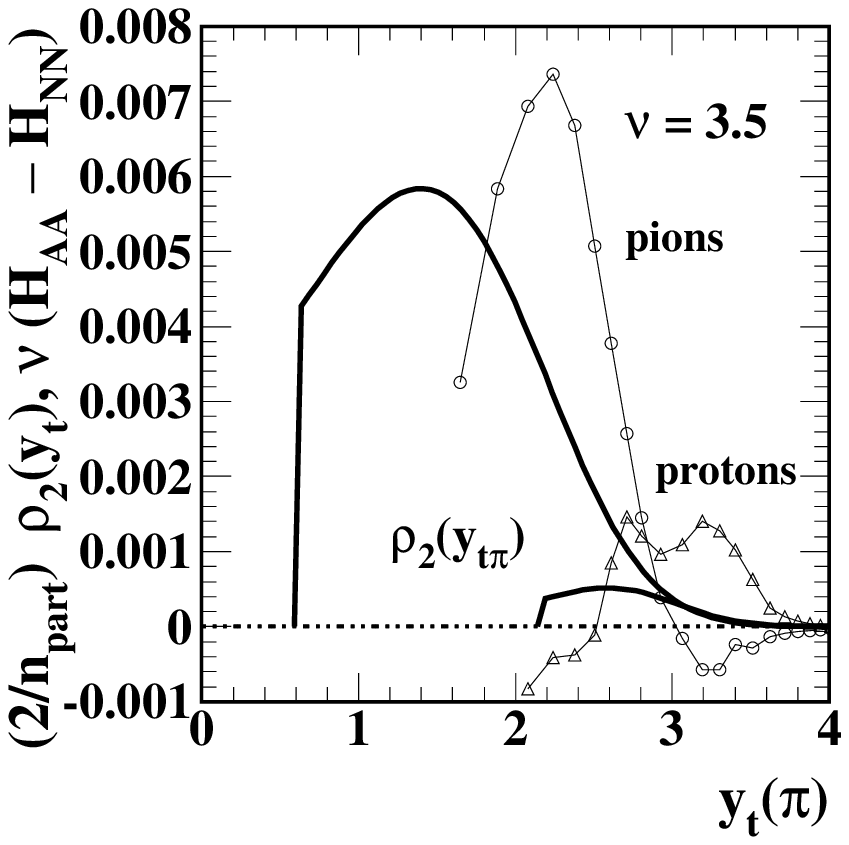}
  \includegraphics[width=1.65in,height=1.65in]{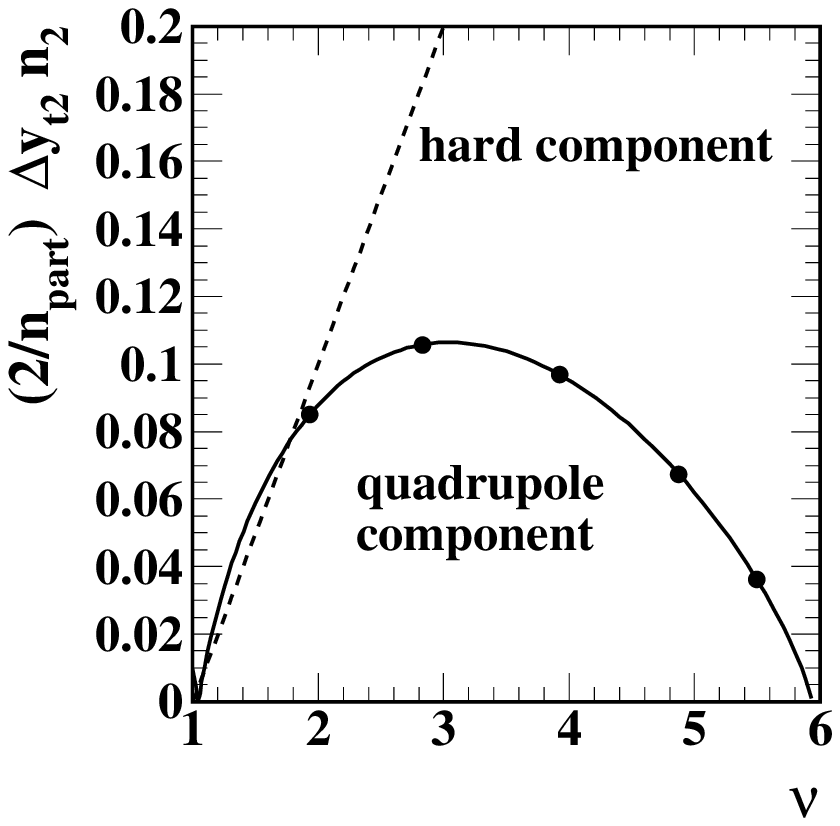}
\caption{\label{aaspectra22}
 Left panel: Minimum values of quadrupole spectra compared to minimum-bias spectrum residuals relative to the two-component model~\cite{2comp}.
Right panel: Expected centrality dependence of the quadrupole integral $2/n_{part}\, \Delta y_{t2}\, n_{ch2}$ compared to that for the spectrum hard component. 
} 
%aaspectra22, aaspectra23
 \end{figure}

Tighter constraints can be established by comparing quadrupole spectrum shapes to {\em residuals} of a comparison between single-particle spectrum data and a two-component spectrum model~\cite{2comp}. Fig.~\ref{aaspectra22} (left panel) shows minimum-bias quadrupole spectra for pions and protons (thick solid curves) compared to the residuals from minimum-bias ($\nu \sim 3.5$) single-particle pion and proton spectra (thin curves and open symbols) obtained by comparing spectrum data with an accurate two-component (soft-plus-hard) model~\cite{2comp}. The proton residuals peak corresponds to the p/$\pi$ ratio ``puzzle"~\cite{friesr,greco,rudy}. The even-larger pion residuals peak was not previously noted. %The proton and pion residuals may also arise from a common boosted source.

%%%%%%%%%%
%%%%%%%%%%

The quadrupole (thick solid) curves have the minimum amplitudes determined above by assuming a positive-definite radial boost $\Delta  y_{t2} \approx \Delta y_{t0}$. The spectrum residuals appear inconsistent with any increase in quadrupole amplitudes beyond the minimum. %Tighter constraints can be established by comparing individual spectrum components and centrality dependencies.

\subsection{Upper limits from centrality dependence}

To confirm the upper limits from minimum-bias spectrum residuals detailed centrality dependence of spectrum structure can be compared on proper hadron rapidity for each species.
%The centrality dependence of the measured $p_t$-integrated quantity $V_2 \sim \Delta y_{t2}\, n_{ch2} \propto \sqrt{\nu}\,  \epsilon\, n_{part}/2$ helps to isolate $n_{ch2}$. 
In Fig.~\ref{aaspectra22} (right panel) the solid curve shows the quadrupole centrality dependence in Eq.~(\ref{quadcent2}) 
\bea \label{quadcent2}
\frac{2}{n_{part}}\, \Delta y_{t2}\, n_{ch2} &\approx& n_{NN} 0.028\, \epsilon \, \sqrt{\nu}
\eea
from the analysis in~\cite{gluequad}. The centrality dependence of the hard component is also sketched for contrast. The minimum-bias $v_2(p_t)$ data used in this analysis correspond to the maximum of product $\Delta  y_{t2} \, n_{ch2}$ on centrality. The product should decrease strongly for more central collisions compared to hard-component structure.

%%%%%%%%%%
 \begin{figure}[h]
  \includegraphics[width=3.3in,height=3.3in]{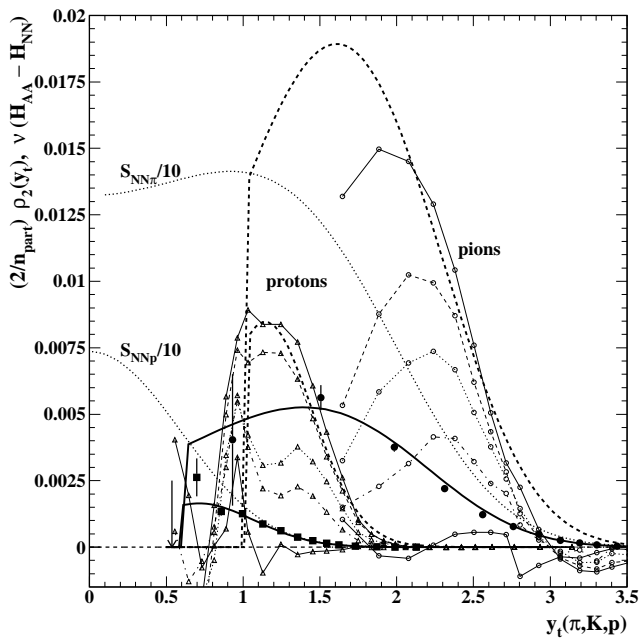}
\caption{\label{aaspectra34}
Lower limits on quadrupole spectra in the form $2/n_{part}\, \rho_2(y_t)$ (solid points and curves) compared to single-particle spectrum residuals (open points and thin curves of various styles) for five Au-Au centralities relative to a two-component reference~\cite{2comp}. Soft components $S_{NN}(y_t)$ provide a reference. The dashed curves corresponding to hadrons from a boosted source with $\Delta y_t \sim 1.1$ and substantially smaller slope parameters $T$ than the quadrupole components suggest a possible mechanism for the residual spectrum structure. 
} 
%aapsectra24
 \end{figure}
%%%%%%%%%%

In Fig.~\ref{aaspectra34} the lower limits of minimum-bias quadrupole spectra for pions and Lambdas (thick solid curves) are compared to the centrality variation of single-particle spectrum residuals on proper hadron $y_t$. The common monopole boost $\Delta y_{t0} = 0.6$ for the pion and Lambda quadrupole curves is apparent. 

Pion and proton spectrum residuals for five centralities are indicated by the thin curves of different types and open symbols. Centralities correspond to the points in Fig.~\ref{aaspectra22} (right panel). Comparison with the boost systematics of the quadrupole spectra suggests that the proton and pion spectrum residuals may also arise from a common boosted source, but with quite different boost distribution. The thick dashed curves sketch a common boosted source with $\Delta y_t \sim 1.1$  for the two hadron species. In that context the boost distributions appear to be strongly centrality dependent. The peak modes move to smaller rapidities for more central collisions. 

The solid points are quadrupole data from Fig.~\ref{boost1ee} with minimum amplitudes as in Fig.~\ref{aaspectra22} (left panel), kinematic factor $p'_t / p_t \gamma_t(1-\beta_t)$ removed and multiplied by $0.025\, T_2 / A \sim 0.5$ according to Eqs.~(\ref{eqxx}) and (\ref{eqyy}). The arrow at left indicates a Lambda $v_2$ datum which is negative but consistent with zero. The agreement between points and solid curves demonstrates that boosted L\'evy distributions describe the quadrupole spectra well.

Comparisons of quadrupole data and spectrum residuals indicate that increasing the quadrupole magnitude beyond the lower limit would strongly conflict with the residuals. The structure and centrality dependence of the proton residuals appear to be consistent with the extrapolated centrality evolution of the lower-limit quadrupole component. The centrality dependence of the structure at the left side of the proton residuals peak may arise from the quadrupole contribution. The pion comparison is indeterminant because of a lack of spectrum data below $y_t \sim 1.7$.  This detailed comparison is limited by sparse data and data uncertainties but suggests that the upper limit on the quadrupole component is consistent with the lower limit. A more precise statement requires improved spectrum and quadrupole data at small $y_t$.

\subsection{Discussion}

The small upper limit ($\sim 5$\%) on the fraction of final-state particles participating in the azimuth quadrupole is certainly counter-intuitive. The conventional scenario for more-central RHIC collisions is that almost all hadrons emerge from a common partially-thermalized medium supporting radial and elliptic flow. The definition of $v_2(p_t)$ implicitly relies on the assumption that the single-particle spectrum in the denominator is the same as the quadrupole spectrum contained in the numerator combined with other factors.

A central message of the present analysis is that the quadrupole and single-particle spectra are not the same, that the former is boosted significantly and the boost is not shared by the single-particle spectrum (radial flow is negligible~\cite{2comp}). Distinctions between spectrum shapes are the basis for estimating what fraction of the final state actually carries the azimuth quadrupole structure.

Since $v_2$ measures the product of the true momentum asymmetry and the fraction of particles carrying the quadrupole, small values of $v_2$ plus the conventional assumption about a flowing bulk medium suggest that the momentum asymmetry is rather small (few percent), and therefore can be explained by hydrodynamic response to initial pressure gradients. 

Comparing the reconstructed quadrupole spectrum with the single-particle spectrum reveals that the quadrupole fraction is actually small, and therefore that the momentum eccentricity for that small fraction of particles is large (near the upper limit defined by a requirement of positive-definite boost). That conclusion is not inconsistent with any previous $v_2$ measurements, only with {\em a priori} expectations within the hydro context.

%%%%%%%%%
\section{Quadrupole {\em vs} nonflow} \label{nonflow}

Fig.~\ref{boost18ab} demonstrates that the hard component of the single-particle spectrum present in the {\em denominator} of $v_2$ severely distorts the structure of $v_2(p_t)$ above about 0.5 GeV/c for pions, kaons and protons. The hard-component distortion should be distinguished from possible {\em nonflow} distortions also due to minijets but appearing in the {\em numerator} of $v_2$. 

Nonflow is dominated by minijet angular correlations misinterpreted as azimuth quadrupole correlations by conventional 1D flow analysis methods~\cite{gluequad}. Minijet correlations and the spectrum hard component have a common source:  hadron fragments from a minimum-bias scattered parton spectrum~\cite{gluequad}.  The combination produces large uncertainties in the interpretation of $v_2$ data above 0.5 GeV/c. In this section I consider the mechanism and consequences of nonflow contributions to $v_2(p_t)$.

\subsection{Nonflow and the hard spectrum component}

The structure of $v_2\{2\}(p_t)$ (1D azimuth correlations) including nonflow is described schematically by
\bea \label{twominis}
v_2\{2\}(p_t,\nu) &\propto& p'_t\, \frac{A\, S'_{NN}(y_t -\Delta y_{_0}) +\text{nonflow}}{S_{NN}(y_t) + \nu H_{AA}(y_t,\nu)}.
\eea
``Nonflow" is dominated by the $m = 2$ azimuth Fourier amplitude of the same-side minijet peak (jet cone) in angular correlations on $(\eta,\phi)$~\cite{gluequad,axialci}. The relative magnitude of the nonflow term in $v_2$ depends in part on the analysis method and spectrum structure. At larger $p_t$ for pions and all $p_t$ for less-abundant hadrons the $v_2\{EP\}$ (event-plane) method is typically employed to accommodate smaller particle yields. $v_2\{EP\} \sim v_2\{2\}$ is maximally sensitive to minijets~\cite{gluequad}. In contrast, 2D angular autocorrelations on ($\eta,\phi$) can be used to separate minijet and quadrupole components accurately~\cite{flowmeth,gluequad}.

Hard component $\nu H_{AA}(y_t,\nu)$ is the angle-integrated minijet fragment spectrum, whereas nonflow is a Fourier component of the minijet same-side peak on azimuth. Thus, minijet contributions in numerator (nonflow) and denominator (hard component) of $v_2(p_t)$ are directly related. However, the nonflow contribution has its own substantial $p_t$ dependence relative to the spectrum hard component (i.e., minijet yield) because the Fourier amplitude of the same-side peak depends on the peak shape ($\eta$ and $\phi$ widths), which varies strongly with parton energy scale (as determined by the selected hadron fragment $p_t$). 

\subsection{${\bf v_2(p_t)}$ trends at larger ${\bf p_t}$}

\begin{figure}[h]
   \includegraphics[width=3.3in,height=1.65in]{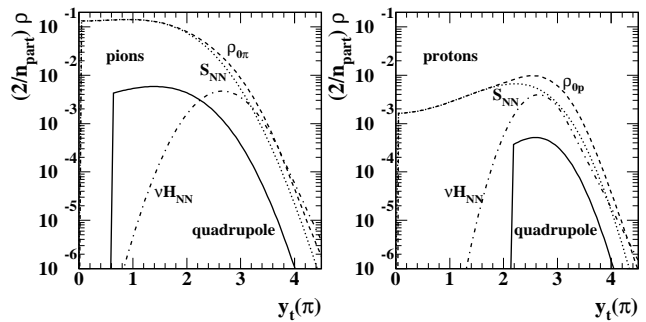}
 \caption{\label{boost6ab}
Comparison of the quadrupole component with soft and hard spectrum components for pions (left panel) and protons (right panel). 
 } 
%boost6ab
 \end{figure}

In Fig.~\ref{boost6ab} quadrupole spectrum components inferred from this analysis are compared to soft and hard single-particle spectrum components for pions and protons from minimum-bias Au-Au collisions. With increasing $p_t$ there is competition between the tails of the quadrupole spectra and hard-component spectra. The latter completely determine $v_2(p_t)$ trends at larger $p_t$. 

%%%%%%%%%%
%%%%%%%%%%

Above $p_t \sim 2$ GeV/c ($y_{t\pi} \sim 3.3$) the soft-component and quadrupole spectra are dominated by the hard component (parton fragments)~\cite{ppprd,2comp}. By definition $v_2\{2\}(p_t)$ represents the $m = 2$ Fourier component of {\em all} azimuth correlation structure. Thus, at larger $p_t$ $v_2\{2\}(p_t)$  is simply the ratio of the Fourier amplitude of the same-side minijet peak (jet cone) to the spectrum hard component (minijets). $v_2(p_t)$ then follows a nearly constant trend on $p_t$ described as ``saturation.'' Variation of the ratio (modulo prefactor $p'_t$) is dominated by changes in the minijet peak shape with parton energy scale. There is no required relation to a reaction plane.

The centrality variation of $v_2(p_t)$ at larger $p_t$ should be dominated by parton energy-loss effects, including $\eta$ broadening of the same-side peak in the numerator (nonflow)~\cite{axialci} and reduction of the hard-component tail in the denominator (``jet quenching'')~\cite{2comp}. Thus, in a conventional flow analysis the region above 0.5 - 1 GeV/c is already strongly distorted by the spectrum hard component and thus difficult to interpret. The region above 3-4 GeV/c provides no information about the quadrupole component, whether hydro or simple boost phenomenon.

%%%%%%%%%%%%%
\section{Quadrupole {\em vs} Scaling} \label{scaling}

Certain {\em scaling relations} are inferred for $v_2(p_t)$ to support claims that ``elliptic flow'' is a hydrodynamic phenomenon manifested by a thermalized bulk partonic medium~\cite{keystone,kolb1,kolb2}, that hadronization from the medium proceeds {\em via} coalescence/recombination of {\em constituent quarks}~\cite{quarkcoal,roy}, relating to a similar model of certain spectrum features (e.g., the anomalous p/$\pi$ ratio at intermediate $p_t$)\cite{fries,greco,rudy}. The overarching conclusion from $v_2(p_t)$ scaling is that sQGP (a thermalized, small-viscosity bulk partonic medium) has been formed. In this section claims of constituent-quark and other forms of $v_2(p_t)$ scaling are re-examined in the context of the present analysis.  

\subsection{$v_2$ scaling observations}

Arguments in favor of a locally-thermalized pre-hadronic bulk medium evolving according to near-ideal hydrodynamics include: 1) the minimum-bias multiplicity distribution form is independent of system size, 2) hadron species abundances follow a statistical model, 3) large $v_2$ values reveal rapid thermalization and early pressure gradients common to all hadron species and incompatible with hydro evolution of hadrons (e.g., $D$ and $\phi$ meson $v_2$ data are interpreted to imply thermalization)~\cite{roy}.

{\em Scaling relations} invoked in flow studies are interpreted to buttress the above arguments. Scaling relations involve combinations of $v_2(p_t)$ data, $m_t$ (transverse mass) and $n_q$ (constituent quark number). The mass dependence of $v_2(p_t)$ at small $p_t$ is attributed to hydrodynamics. Constituent quark scaling expressed by $v_2^h(p_t) = n_q\, v_2^q(n_q\, p_t^q)$, with $n_q = 2$ for mesons and 3 for baryons~\cite{quarkcoal} is interpreted to imply hadronization from a thermalized partonic medium. 

In~\cite{roy} a universal scaling of the combination $v_2(p_t) / \epsilon\, n_q$ {\em vs} ``kinetic energy'' $(m_t - m_0)/n_q$ was claimed over a broad range of centralities, strongly suggesting formation of a thermalized partonic medium. However, other measurements disagree with the claimed universal centrality trend~\cite{v2over}. Universal scaling results are also claimed for 30-70\% centrality, but one can ask when is the system not in equilibrium? For what circumstances do such scaling trends not hold? What collision systems (e.g., N-N) do not thermalize or form a ``perfect liquid?'' 

The present analysis strongly suggests that most hadrons emerge from several nearly-independent QCD processes (nucleon or parton scattering and fragmentation), but some coupling develops among the processes in more-central Au-Au collisions. In Sec.~\ref{structure} it was shown that there are two shape factors in $v_2(p_t)$: $p'_t$ in the boost frame and the spectrum ratio $S'_{NN}(y_t - \Delta y_{t0}) / \rho_0(y_t)$. In what follows I consider scaling arguments for each factor.

\subsection{$v_2$ scaling and ${\bf p'_t}$}

In Fig.~\ref{boost3cbx} (right panel) $p'_t$ {\em vs} $p_t$ is plotted. $\gamma_t\, (1 - \beta_t) \sim 0.6$ common to three hadron species determines all structure. Similar ``mass scaling'' of $v_2(p_t)$ is taken to imply hydrodynamic flow. But the mass dependence near the origin is determined by a single radial (monopole) boost $\Delta y_{t0}$, and there is no indication from such data of the actual boost and hadron production mechanisms.

%%%%%%%%%%
 \begin{figure}[h]
   \includegraphics[width=1.65in,height=1.65in]{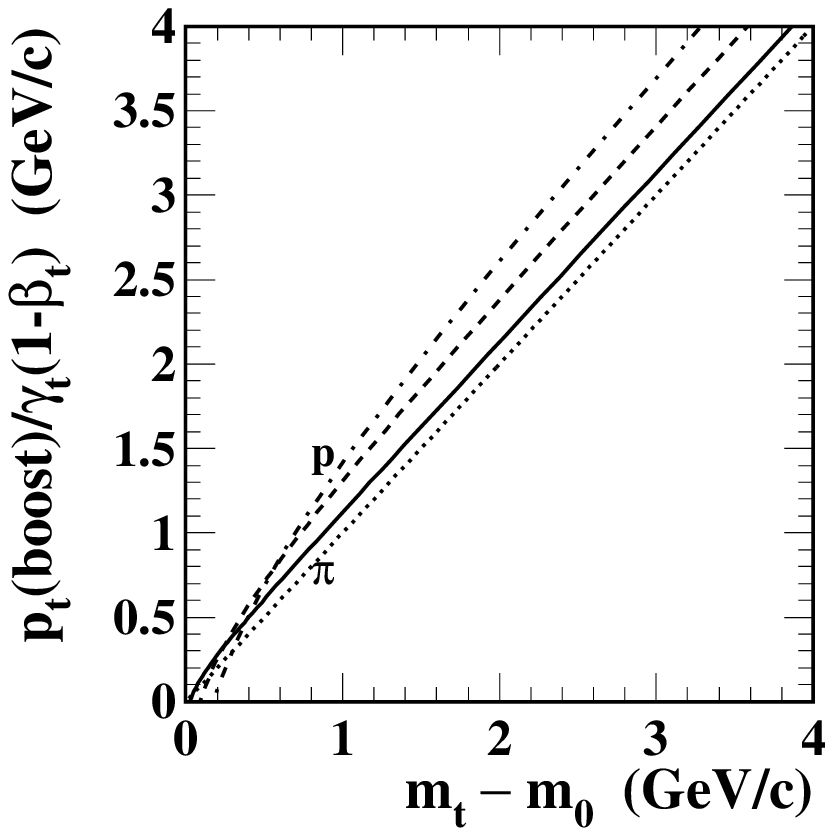}
   \includegraphics[width=1.65in,height=1.65in]{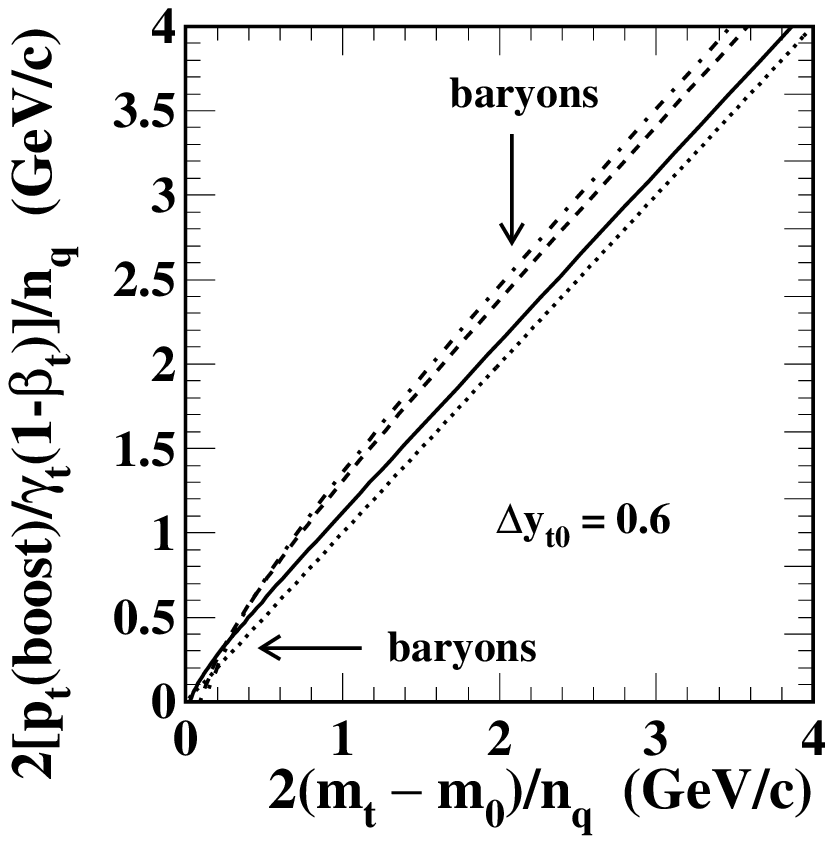}
\caption{\label{boost3be}
Left panel:   $p_t$ in the boost frame compared to transverse kinetic energy $m_t - m_0$ in the lab frame for monopole boost value $\Delta y_{t0} = 0.6$.  The curve intercepts are at $m_{t0} - m_0$ as defined in the text.
Right panel: The same relations rescaled by ``constituent quark number'' as $2/n_q$ so that meson trends are unchanged. The shift for baryons is indicated.
 } 
%boost3b, boost3e
 \end{figure}
%%%%%%%%%%

In Fig.~\ref{boost3be} (left panel) $p'_t$ is replotted on $m_t - m_0$. The mass dependence near the origin {\em appears} to be reduced, but the locations of the curve intercepts are simply given by $m_{t0} - m_0 =  m_0(\cosh[\Delta y_{t0}]-1)\sim  m_0 \, (\Delta y_{t0})^2/2 $ on $m_t - m_0$ compared to $p_{t0} = m_0\, \sinh(\Delta y_{t0})\sim m_0 \Delta y_{t0}$ on $p_t$ noted in Fig.~\ref{boost3} (left panel). Consequences of the source boost, especially boost distribution details, are compressed on $m_t - m_0$ (by a factor 3 for $\Delta y_{t0} \sim 0.6$) in the $p_t$ region {\em most important to the hydro interpretation}, but the boost is just as accurately determined from the data regardless of plotting format. Fig.~\ref{boost1a} (right panel) clearly provides the best visual access. At larger $p_t$ $p'_t \rightarrow m_t$ is shifted upward by $m_0$ relative to abscissa $m_t - m_0$.

In Fig.~\ref{boost3be} (right panel) both axes are scaled by $2/n_q$ (factor 2 so axis values remain the same for comparison). The consequences are trivial. The intercept at smaller $m_t$ is reduced by 2/3 for baryons, and the constant vertical offset $m_0$ at larger $m_t$ is also reduced by 2/3 for baryons. Visual differences between baryons and mesons are indeed reduced, but the {results are not fundamental} because the form of $v_2(p_t)$ at larger $p_t$ is {\em not determined by hydro or any boost phenomenon} (cf. next subsection).

\subsection{$v_2$ scaling and spectrum ratios}

\begin{figure}[h]
   \includegraphics[width=1.65in,height=1.65in]{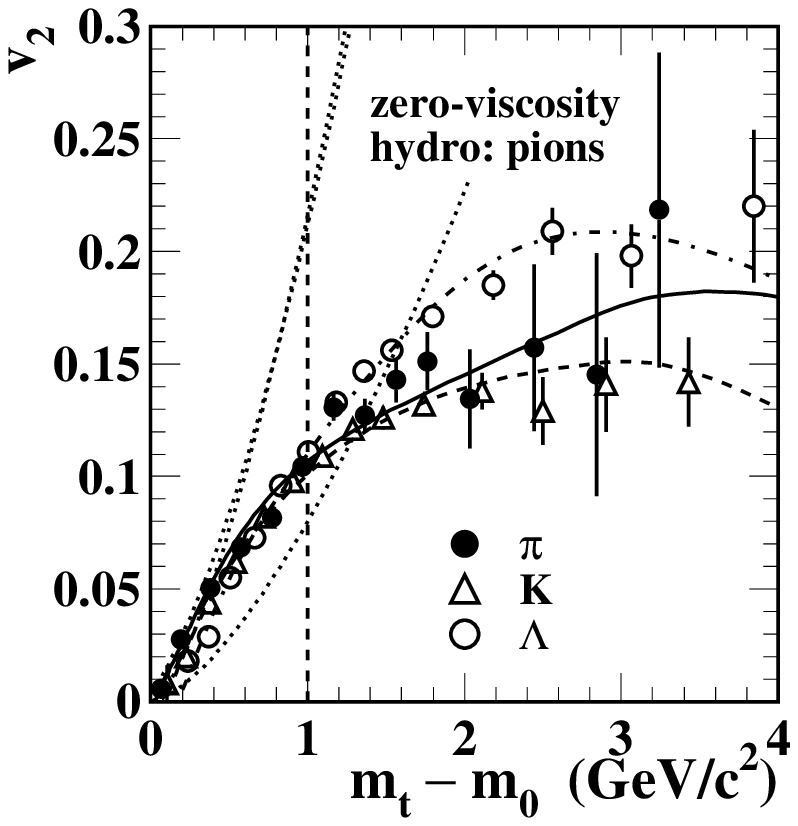}
   \includegraphics[width=1.65in,height=1.65in]{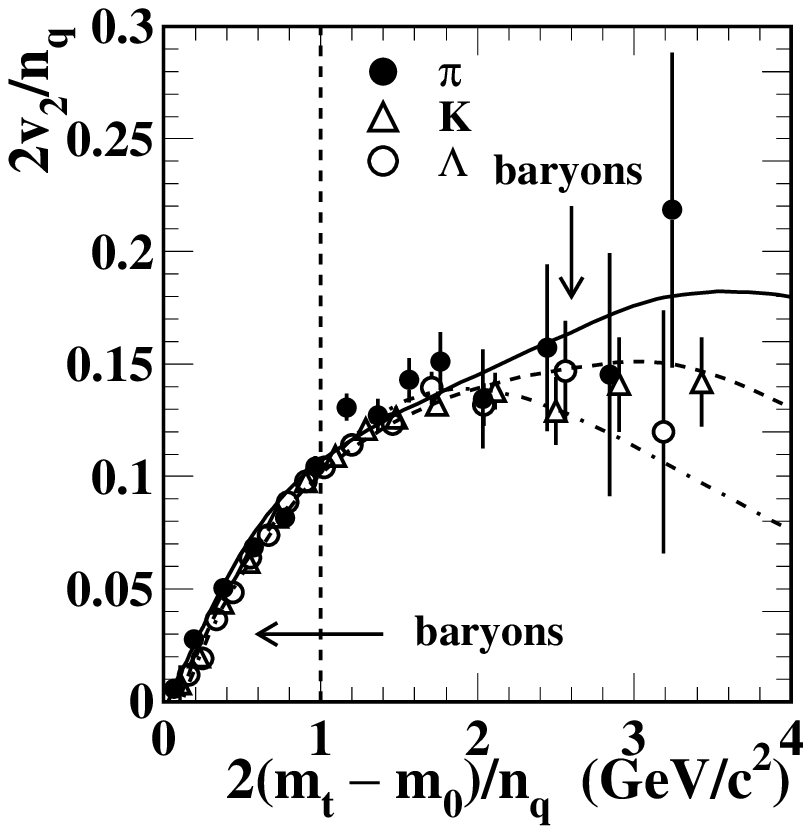}
\caption{\label{boost1fg}
Left panel:  $v_2(p_t)$ is plotted {\em vs} kinetic energy $m_t - m_0$. The intercept spacing at small $p_t$ is reduced by a factor 3$\times$. The lower dotted curve is the zero-viscosity hydro curve~\cite{teaney}. The upper dotted curve is the hydro curve $\times 2.7$.
Right panel: The left panel except with constituent quark scaling in the form $2/n_q$ so that meson trends remain unchanged. The shift for baryons is indicated by the arrows.
 } 
%boost1f, boost1g
 \end{figure}

Fig.~\ref{boost1fg} (left panel) shows $v_2(p_t)$ data for three hadron species plotted in the same format as Fig.~\ref{boost3be} (left panel). This figure can be compared directly with Fig.~4 of~\cite{roy}. The dotted curves represent the hydro theory curve~\cite{teaney} and $2.7\times$ hydro. Data near the origin follow the $p'_t$ systematics described above. As in Sec.~\ref{specrat} (spectrum ratios) the turnover of $v_2(p_t)$ above 0.5 GeV/c is due to the hard component in the $v_2$ denominator. If $v_2$ data do not return to zero at larger $m_t$ a significant nonflow contribution is probably present, as discussed in Sec.~\ref{nonflow}.

%%%%%%%%%%
%%%%%%%%%%

In Fig.~\ref{boost1fg} (right panel) the $n_q$ scaling strategy is used to minimize apparent differences between baryon and meson data, the resulting shifts indicated by the arrows. The most dramatic changes occur above 1 GeV/c where the data are not relevant to a hydrodynamic mechanism or soft processes. In scaling exercises the region above 1 GeV/c is viewed as dominated by elliptic flow and soft hadronization. Scaling trends there are interpreted in turn to imply that hadron production is dominated by coalescence of constituent quarks. 

The apparent correspondence of data for different hadron species left of the vertical dashed lines indicates that all-important information about the source boost distribution (cf. Fig.~\ref{boost18cd} -- right panel) has been made {\em visually inaccessible} by a simple transformation. Note the limited region of comparison between the upper hydro (dotted) curve and the data. Generally, comparison of different hadron species on $p_t$ or $m_t$ rather than $y_t$ is unsuited for hydro (common boost) phenomena. 

$v_2$ data to the right of the vertical dashed lines can reveal nothing about hydrodynamic phenomena. The $v_2(p_t)$ ratio there is dominated by a complex mixture of hard processes (parton scattering and fragmentation), soft spectrum components and quadrupole components, with different shape parameters ($T,n$, etc.) for each component. Sec.~\ref{specrat} reveals that the systematics of Fig.~\ref{boost1fg} are determined by the mass dependence of several spectrum components reflecting soft and hard processes. 
%Deviation of various hadron species from a constant trend, which {\em must occur} at larger $p_t$ absent minijet contamination, are suppressed in plots terminated at $p_t = 2-3$ GeV/c.

The present analysis demonstrates that quadrupole spectra are similar to soft spectrum components (L\'evy distributions) unchanged from N-N collisions. The quadrupole hadron production mechanism may well be the same as in elementary collisions. Inference of ``constituent quark scaling'' from $v_2$ data is prompted by a combination of several conventional collision mechanisms confused by a poorly-designed correlation measure.

%%%%%%%%%
\section{Discussion}

\subsection{The conceptual context of elliptic flow}

%Elliptic flow ($v_2$) is interpreted in the context of a thermalized partonic bulk medium. 
The conventional elliptic flow context is a limiting case in which $v_2$ measurements are interpreted to conclude that 1) a monolithic bulk medium produced early in the collision  is ``partonic'' (QCD quanta dominate the dynamics), 2) the medium is thermalized rapidly {\em via} partonic rescattering and 3) hadrons emerge late from the medium {\em via} coalescence of constituent quarks. Multistrange $v_2$ data for instance exclude slower flow development via hadron thermalization (rescattering)~\cite{v2strange}.

Spectrum and elliptic flow systematics are used to conclude that ``constituent quarks'' play a role in hadronization. Recombination (quark coalescence)~\cite{quarkcoal,rudy,fries,greco} reproduces ``many features'' of hadron spectra in the intermediate $p_t$ region [1.5,5] GeV/c according to~\cite{0701010}. Hadronization by quark coalescence is also inferred from ``scaling'' of $v_2(p_t)$ at intermediate $p_t$, where $v_2$ is said to ``saturate'' at a number apparently $\propto n_q$ above 1 GeV/c~\cite{0701010}.  {\em Limiting values} of $v_2$ (i.e., hydro limits) combined with quark-number scaling (dynamical DoF are constituent quarks) ``suggest that strongly-coupled matter with sub-hadronic degrees of freedom may be created in heavy ion collisions at RHIC''~\cite{0701010}. The present analysis is inconsistent with those conclusions.

\subsection{The fragmentation alternative}

The  complementary limiting case is A-A collisions modeled by {\em linear superposition} of N-N collisions according to the Glauber model, and hadron production by {\em in vacuo} nucleon and parton fragmentation as in elementary hadronic collisions. That {\em linear reference} invokes two independent fragmentation processes to describe A-A collisions: participant-nucleon (soft, longitudinal) fragmentation leading to a {\em soft component} of $p_t$ spectra and minimum-bias large-angle-scattered parton (hard, transverse) fragmentation leading to a {\em hard component}. 

Some aspects of fragmentation can {\em appear} thermal, even described in part by the statistical model, although there is no transport via binary collisions (rescattering) in the Boltzmann sense. Fragmentation is a maximum-entropy process, the entropy maximization achieved via splitting cascades. Parton fragmentation in LEP $e^+$-$e^-$ collisions is described to the statistical limits of data by the beta distribution, a maximum-entropy function~\cite{lepmini}. Deviations from the linear-superposition model in more-central A-A collisions  could result from a few secondary parton interactions. Any non-linearities require {\em careful differential study} relative to a linear reference (the two-component model)~\cite{ppprd,2comp}. The burden should be on claims of thermalization to {\em rule out} fragmentation as the dominant mechanism of A-A collisions at RHIC.

\subsection{Importance of measure design}

Ratio measures applied to RHIC data typically confuse several collision mechanisms. The $R_{AA}$ spectrum ratio mixes soft and hard spectrum components. Most of the hard component (at smaller $p_t$) is obliterated by the soft component in a measure intended specifically to study parton energy loss~\cite{2comp}. Ratio $v_2(p_t)$ similarly mixes soft and hard spectrum components. This analysis demonstrates that the (soft) quadrupole component (boosted source) is severely distorted by the hard component (parton fragmentation) over most of the $p_t$ acceptance. The benefits of improved measure design are suggested by comparison of Fig.~\ref{boost1a} (left panel) with Fig.~\ref{boost1ee}.

Some conventional single-particle spectrum analysis also produces misconceptions. The results of monolithic ``power-law'' function~\cite{ua1} fits to multicomponent $p_t$ spectra cannot be interpreted~\cite{ppprd}. If the entire spectrum below some $p_t$ value (e.g., 2 GeV/c) is described by a blast-wave model~\cite{blastwave} the abundant hard spectrum component (minijets) is injected into the hydro parameterization, confusing parton fragmentation with hydrodynamic (Hubble) expansion~\cite{2comp}. Better understanding of RHIC collisions requires a comprehensive differential approach to single-particle and correlation measurements, including comparisons to well-defined references. 

\subsection{Comparison of boost models}

In Sec.~\ref{boostc} two radial boost models were described and model 2, multiple hadron sources including a boosted quadrupole component, was adopted for the present analysis. Model 1 is the conventional thermalized partonic bulk medium common to all soft hadrons. I reconsider the model choice in light of the analysis results.

Model 1 is essentially the {\em blast-wave} model of heavy ion collisions~\cite{starblast} applied by hypothesis to {\em almost all particle production}. Uniform Hubble expansion is assumed for longitudinal and radial boosts. The longitudinal system is boost invariant; the transverse boost depends on radius (Hubble expansion) and azimuth (elliptic flow). Particle emission angle $\phi_p$ is distinguished from particle source azimuth $\phi_s$ and the normal to the emission surface $\phi_b$. The source pseudorapidity $\eta$ (polar angle) is not generally the same as the particle longitudinal rapidity $y_z$.

In the present analysis the soft component, quadrupole component and hard component are decoupled. I model the quadrupole component by normal emission from a cylinder at mid-rapidity and $z = 0$. The blast-wave model simplifies to $y_z = \eta = 0$ and $\phi_s = \phi_p = \phi_b$. Eq.~(11) of~\cite{starblast} then becomes the first line of Eq.~(\ref{boostkine}) of this paper. Since this analysis emphasizes qualitative study of algebraic structure the simplifications are reasonable. The boost model of~\cite{starblast} includes $\rho_0 \rightarrow \Delta y_{t0}$ and $\rho_2 \rightarrow \Delta y_{t2}$. However, parameters $\rho$ relate to a Hubble expansion model whereas the $\Delta y_t$ relate to an expanding cylindrical shell. $\Delta y_{t0} \sim 0.6\, \rho_0$ and similarly for the quadrupole. Model 1 is thus {\em a limiting case} of model 2.

The model-1 expectation is that $T$ and $\rho_0$ common to most particle production are obtained from $p_t$ spectrum fits, and $\rho_2$ is obtained from fits to $v_2$ data. The result in~\cite{starblast} for the monopole component is $\rho_0 \sim 0.9$ or $\Delta y_{t0} \sim 0.55$ independent of centrality. Close inspection of spectrum fits however reveals that the description of the overall spectra over the full $p_t$ range is poor, especially for pions (the model is fitted to data over a very restricted $p_t$ interval). Compare with the detailed spectrum description in~\cite{2comp} in the $p_t$ interval [0.2,12] GeV/c where no radial boost was required. Model-1 attribution of a transverse boost system to the {\em majority} of particles appears to fail, consistent with the present analysis.

The approximate quadrupole spectra determined by data points in Fig.~\ref{boost1ee} are obtained from a simple combination of measured quantities motivated by Eq.~(\ref{stuff}). The common boost $\Delta y_{t0}$ of the quadrupole spectra is not observed for single-particle soft components---the monolithic boost distribution of the blast-wave model is inconsistent with data. The sharp edge of the quadrupole spectrum (narrow boost distribution) is particularly inconsistent with Hubble expansion, as shown in Fig.~\ref{boost18cd}. Independence of the quadrupole component boost from the soft component is thus in conflict with model 1.

\subsection{Implications of the present analysis}

\begin{itemize}

\item  Hadron $y_t$ spectra associated with the quadrupole component have been recovered from $v_2(p_t)$ data

\item $v_2(p_t)$ data trends are revealed as a complex interplay of three hadron production mechanisms with {\em accidental} manifestations of mass dependence

\item The structure of the $v_2(p_t)$ {\em ratio} is dominated by the spectrum hard component above 0.5 GeV/c

\item Quadrupole hadrons come from a boosted source with narrow boost distribution not common to most hadrons; the number of hadrons from the quadrupole source is a small fraction of the total

\item The small-$p_t$ mass ordering invoked to support a hydro interpretation is a kinematic consequence of any common boosted source

\item The quadrupole component appears to be isolated from the rest of the collision evolution

\item Quadrupole spectra are substantially ``cooler'' than the single-particle spectrum soft component (i.e., the quadrupole and soft components are {\em not} in thermal equilibrium, with each other or with the spectrum hard component)

\item $v_2(p_t)$ trends interpreted as ``constituent quark scaling'' at intermediate $p_t$ do not relate to a hydro phenomenon or to hadron formation from a thermalized partonic medium

\end{itemize}

%%%%%%%%%
 \section{Summary}

Elliptic flow ($v_2$) measurements provide the primary support for claims of ``perfect liquid'' at RHIC. That central role motivates a careful re-examination of the interpretation of $v_2$ data in terms of hydrodynamic models. To that end I have reviewed azimuth correlation analysis methods and provided important generalizations. I described a method to extract {\em quadrupole spectra} on $y_t$ from $v_2(p_t)$ data, and I used a limited $v_2$ data sample for identified hadrons from minimum-bias Au-Au collisions at 200 GeV to illustrate properties of $v_2$ inferred from $p_t$ dependence and mass dependence.

I reviewed an accurate two-component parameterization of hadron single-particle spectra on $y_t$ required to extract quadrupole spectra from $v_2(p_t)$, and introduced the kinematics of boosted sources as an aid to interpreting features of $v_2(p_t)$. I expressed the functional form of $v_2(p_t)$ as the product of two factors: $p'_t$ ($p_t$ in a boosted frame) and the ratio of the sought-after quadrupole spectrum to the single-particle spectrum 

I described the analysis steps required to combine the above elements so as to recover quadrupole spectra from $v_2(p_t)$ data and modeled the extracted quadrupole spectra with L\'evy distributions---boosted soft components $S'_{NN}(y_t - \Delta y_{t0})$. I compared the quadrupole spectrum component quantitatively to other spectrum components and to two hydro theory examples, and I estimated absolute quadrupole yields. Finally, I considered the impact of nonflow (minijet) contributions to $v_2$ measurements

The conclusions from this analysis are as follows: Claims for $v_2$ {\em scaling behavior} supporting inference of a major role for constituent quarks in collision dynamics appear to be unsupported given the structure of  ratio $v_2(p_t)$ and mixing of different physical mechanisms by that measure, especially above $p_t \sim 0.5$ GeV/c. The true universality, as in Fig.~\ref{boost1a} (right panel), is that of hadrons emitted from a common boosted source by the same hadronization mechanism as the single-particle spectrum soft component, albeit with a smaller ``temperature.'' There is no support for a novel hadron production mechanism. Monopole boost $\Delta y_{t0}$ is accurately obtained from $v_2(p_t)$ data, but the (small) quadrupole absolute yield is inferred indirectly, since only the product of quadrupole boost $\Delta y_{t2}$ and absolute yield $n_{ch2}$ is measured directly. 

Analysis of data for three hadron species indicates that quadrupole yields {\em relative to} spectrum soft components are similar. The production mechanism for the soft-component yields in N-N collisions is the mechanism for the quadrupole yields in Au-Au collisions. Thus, {\em only three numbers} (two boosts and a ratio) are obtained from minimum-bias $v_2(p_t)$ data. Quadrupole ($v_2$) data provide no evidence for a thermalized system or for medium properties such as viscosity. The quadrupole component appears to result from an isolated dynamical process involving at most $5$\% of the hadrons in Au-Au collisions. 

The combination of those properties suggests that the azimuth quadrupole may be a new QCD phenomenon emerging at smaller QCD energy scales, interaction of QCD fields over large space-time volumes, which does not couple significantly to other collision processes and produces a hadron spectrum significantly ``cooler'' than the spectrum soft component from nucleon fragmentation. The smaller slope parameter may result from reduced $k_t$ broadening of the QCD quadrupole field component compared to nucleon fragmentation (soft component).

This work was supported in part by the Office of Science of the U.S. DoE under grant DE-FG03-97ER41020.

%%%%%%%%%%%%%%%%%%%%%%%%%%

%%%%%%%%%%%%%%%%%%%%%%%%%%%%

\end{document}